\documentclass[aps,preprint,nofootinbib,floatfix,prd]{revtex4-1}
\usepackage{color}
\usepackage[dvipsnames]{xcolor}

\pdfoutput=1
\usepackage{graphicx}
\usepackage[latin1]{inputenc}
\usepackage{amsmath,amssymb}
\usepackage{slashed}
\usepackage{hyperref}
\usepackage{epstopdf}

\newcommand{\lsim}{\mathrel{\mathop{\kern 0pt \rlap
  {\raise.2ex\hbox{$<$}}}
  \lower.9ex\hbox{\kern-.190em $\sim$}}}
\newcommand{\gsim}{\mathrel{\mathop{\kern 0pt \rlap
  {\raise.2ex\hbox{$>$}}}
  \lower.9ex\hbox{\kern-.190em $\sim$}}}

\def  \bcen   {\begin{center}}
\def  \ecen   {\end{center}}
\def  \beq    {\begin{equation}}
\def  \eeq    {\end{equation}}
\def  \bpm    {\begin{pmatrix}}
\def  \epm    {\end{pmatrix}}
\def  \beqa   {\begin{eqnarray}}
\def  \eeqa   {\end{eqnarray}}
\def  \nn     {\nonumber }
\def\bea{\begin{eqnarray}}
\def\eea{\end{eqnarray}}

\def\eps {\epsilon}
\def\ga   {\gamma}
\def\Ga   {\Gamma}
\def\th   {\theta}
\def\la   {\lambda}
\def\La   {\Lambda}
\def\al   {\alpha}
\def\sig   {\sigma}

\def\nn{\nonumber}
\def\lee { \left( }
\def\rii { \right) }
\def\lan   {\langle}
\def\ran   {\rangle}

\def\De {\Delta}
\def\ka  {\kappa}

\def\to {\rightarrow}

\def\tot {\leftrightarrow}

\begin{document}

{\small
\begin{flushright}
CP3-Origins-2018-026 DNRF90
\end{flushright} }

\title{
Circularly Polarized Gamma Rays\\ 
in Effective Dark Matter Theory
}
\author{
Wei-Chih Huang$^1$, Kin-Wang Ng$^{2,3}$ and Tzu-Chiang Yuan$^2$
}

\affiliation{
\small{
$^1$CP$^{\, 3}$-Origins, University of Southern Denmark, Campusvej 55, DK-5230 Odense M, Denmark \\
$^2$Institute of Physics, Academia Sinica, Nangang, Taipei 11529, Taiwan \\
$^3$Institute of Astronomy and Astrophysics, Academia Sinica, Nangang, Taipei 11529, Taiwan
}
}

\date{\today}

\begin{abstract}

We study the loop-induced circularly polarized gamma rays from dark matter annihilation using the effective dark matter theory approach.
Both neutral scalar and fermionic dark matter annihilating into monochromatic diphoton and $Z$-photon final states are considered. 
To generate the circular polarization asymmetry, $P$ and $CP$ symmetries must be violated in the couplings 
between dark matter and Standard Model fermions inside the loop with non-vanishing Cutkosky cut.
The asymmetry can be sizable especially for $Z$-photon final state for which asymmetry of nearly $90\%$ can be reached. 
We discuss the prospect for detecting the circular polarization asymmetry of the gamma-ray flux from dark matter annihilation in the Galactic Center in future gamma-ray polarimetry experiments.

\end{abstract}

\maketitle

\section{Introduction} \label{section:intro}

Recent cosmological observations such as the cosmic microwave background, large scale structures, and galactic rotational curves concordantly imply the existence of a large amount of cold dark matter (CDM) in the Universe~\cite{Ade:2015xua,Aghanim:2018eyx}.
The most well-motivated candidate for the CDM is a non-Standard Model (SM) particle such as the lightest supersymmetric particle, extra dimension, hidden sector, and Higgs portal dark matter (DM). So far all experimental searches for these particles remain elusive, giving us stringent constraints to their interactions with the SM sector (see Ref.~\cite{Tanabashi:2018oca} for a recent review).

Unlike the hunting for DM signals as missing energies in colliders or the direct measurement in cryogenic detectors of the recoil energy of target nuclei bombarded by Galactic halo DM particles, the indirect observation of DM particles accumulated in the Solar core or Galactic Center via their decay or annihilation products such as gamma~($\ga$) rays, positrons, antiprotons, and neutrinos, usually suffers from uncontrollable astrophysical environment and background. There are claims from time to time of excess diffuse or Galactic $\ga$ rays and excess positron flux in cosmic rays that are above the astrophysical background, but interpreting the data as the DM signal may be wrong without a thorough removal of the astrophysical uncertainties. As such, any distinct feature of the potential DM signal that helps distinguish DM particles from astrophysical sources would be very invaluable.
The capability of detecting polarization in future $\ga$-ray observations will provide 
a new tool to probe the nature of DM~\cite{Aharonian:2017xup}. 
It is certainly important to explore the possibility for a net linear or circular polarization of $\ga$ rays coming from 
DM annihilations or decays. Recent work along this direction can be found in 
Refs.~\cite{Ibarra:2016fco,Kumar:2016cum,Gorbunov:2016zxf,Boehm:2017nrl,Elagin:2017cgu,Huang:2018qui,Queiroz:2018utk}. 
Most of these studies are based on specific model buildings.
In this {\it Letter}, we will study the circular polarization of $\ga$ rays from DM annihilations using the effective DM theory approach. 
This is a systematic and model independent way of studying the annihilation of neutral DM particles into $\ga$ rays 
that allows us to obtain the conditions for a net circular polarization of the $\ga$ rays.

In the following, we start with the effective operators of interest for both scalar and fermion DM in Sec.~\ref{section:DMEFT}.
A little digression addressing the convention of photon polarization as well as fermion spinor wavefunction 
is devoted in Sec.~\ref{sec:polar_spin}. 
We then proceed to compute loop-induced  amplitudes of different photon polarizations for the diphoton final state, followed by the $Z\ga$ channel
in Secs.~\ref{sec:diphoton} and \ref{sec:Z_gamma} respectively.
The numerical results are presented in Sec.~\ref{sec:results}, and the summary and prospects of future observations are discussed in Sec.~\ref{sec:conclusions}.

\section{Effective Dark Matter Interactions} \label{section:DMEFT}

For complex scalar DM, we consider the following dimension 5 effective operator
\begin{equation}
\label{eq:effL_S}
{\mathcal L}^S = \frac{(4 \pi)^2}{\Lambda} 
\left(  \chi^* \chi  \right) \left( \bar f \left( C_f^S + i C_f^P \gamma_5 \right) f \right) \; ,
\end{equation}
where $C_f^S$ and $C_f^P$ are real and $f$ refers to a SM charged fermion.
$\Lambda$ is an effective high cutoff scale which we do not 
need to specify explicitly.

For Dirac DM, we will focus on three dimension 6 effective operators. The first one is
\begin{equation}
\label{eq:effL_D1}
{\mathcal L}^D_1 = \frac{(4 \pi)^2}{\Lambda^2} 
\left(  \bar \chi \left( C_\chi^S + i C_\chi^P \gamma_5 \right) \chi  \right) 
\left( \bar f \left( C_f^S + i C_f^P \gamma_5 \right) f \right) \; ,
\end{equation}
with real $C_{f,\chi}^S$ and $C_{f,\chi}^P$. The second one is
\begin{equation}
\label{eq:effL_D2}
{\mathcal L}^D_2 = \frac{(4 \pi)^2}{\Lambda^2} 
\left(  \bar \chi \gamma_\mu \left( C_\chi^L P_L + C_\chi^R P_R \right) \chi  \right) 
\left( \bar f \gamma^\mu \left( C_f^L P_L + C_f^R P_R \right) f \right) \; .
\end{equation}
Since the coefficients $C_{\chi , f}^{L,R}$ are necessarily real, there is no complex parameter in $\mathcal L^D_2$, 
we do not expect it will give rise to net circular polarization. 
This will be checked by explicit calculations in Secs.~\ref{sec:diphoton} and \ref{sec:Z_gamma}.
The last one is
\begin{equation}
\label{eq:effL_D3a}
{\mathcal L}^D_3 = \frac{(4 \pi)^2}{\Lambda^2} 
\left(  \bar \chi \sigma_{\mu \nu} \left( \widetilde C_\chi^{L} P_L + \widetilde C_\chi^{R} P_R \right) \chi  \right) 
\left( \bar f \sigma^{\mu \nu}  \left( \widetilde C_f^{L} P_L + \widetilde C_f^{R} P_R \right) f \right) \; ,
\end{equation}
with $\widetilde C_\chi^R = ( \widetilde C_\chi^L )^*$ and $\widetilde C_f^R = ( \widetilde C_f^L )^*$.
Note that we have normalized the effective operators in Eqs.~\eqref{eq:effL_S}-\eqref{eq:effL_D3a} 
according to the naive dimensional analysis~\cite{Manohar:1983md}. 

Using the self-duality identities
\begin{equation}
\frac{i}{2} \epsilon^{\alpha\beta\mu\nu} \sigma_{\mu \nu}P_{L,R} = \pm \sigma^{\alpha\beta}P_{L,R} \, ,
\end{equation}
$\mathcal L^D_3$ in Eq.~\eqref{eq:effL_D3a} can be rewritten as
\begin{equation}
\label{eq:effL_D3b}
{\mathcal L}^D_3 = \frac{(4 \pi)^2}{\Lambda^2} \left[
\widetilde C_{\chi f}  \left(  \bar \chi_R \sigma_{\mu \nu}  \chi_L  \right) \left( \bar f_R \sigma^{\mu \nu}   f_L \right) 
 +
\widetilde C_{\chi f}^* \left(  \bar \chi_L \sigma_{\mu \nu}  \chi_R  \right) \left( \bar f_L \sigma^{\mu \nu}   f_R \right) \right] \; ,
\end{equation}
with $\widetilde C_{\chi f} = \widetilde C_\chi^{L}  \widetilde C_f^{L}$ 
and $\widetilde C_{\chi f}^* =  \widetilde C_\chi^{R}  \widetilde C_f^{R} 
= (\widetilde C_\chi^{L}  \widetilde C_f^{L})^*$.

It is straightforward to extend our analysis to the case of real scalar or Majorana DM.
For the Majorana case, we note that  $\bar \chi \gamma_\mu \chi = 0$, 
$\bar \chi \sigma_{\mu \nu}  \chi = 0$, $\bar \chi \sigma_{\mu \nu}  \gamma_5 \chi = 0$.
We will not consider these cases any further here.

\section{Photon Polarization Vectors and Fermion Spinors}~\label{sec:polar_spin}

In this section, we specify the convention on photon polarization and fermion spinor wavefunction
adopted in this work. 
Following Refs.~\cite{Hagiwara:1985yu,Boehm:2017nrl},
a photon traveling with 4-momenta $k^\mu=(k_0, \, 0, \, 0, \, k_z)$ 
has the right-handed ($+$) and left-handed ($-$) circular polarization vectors denoted by
\begin{align}
\epsilon^\mu_\pm =\frac{1}{\sqrt{2}} \left( \mp \epsilon^\mu_1 - i \epsilon^\mu_2 \right) \, ,
\label{eq:pol_vec}
\end{align} 
with 
\begin{align}
\epsilon^\mu_1 (k)=\left( 0, \, 1, \, 0, \, 0 \right) \;\; , \;\;
\epsilon^\mu_2 (k)=\left( 0, \, 0, \,  \frac{k_z}{ \vert k_z \vert} , \, 0 \right) \;\; . \;\;
\label{eq:pol_vec_1}
\end{align}
For the $Z$ boson, besides the above two circular polarization vectors, we also need
\begin{align}
\epsilon^\mu_L (k)=\frac{1}{m_Z}\left( k, \, 0, \, 0, \, \sqrt{k^2 + m^2_Z} \right) \;\; , \;\;
\end{align}
for the longitudinal component of the $Z$ boson with 3-momentum ${\bf k} = k\hat{z}$ along the $+ \hat{z}$ direction
as we will discuss the $Z\ga$ final state as well. 

On the other hand, assuming the Dirac DM pair annihilates at rest,
the four-component spinor for a Dirac DM particle $\chi$ of spin $s=1/2$ in the limit of $v_{\text{DM}} \to 0$
is
\begin{align}
u^s(|\vec{p}|=0) =\sqrt{m_\chi}
\begin{pmatrix}
 \xi^s \\ 
\xi^s
\end{pmatrix} \, ,
\label{eq:u_wav}
\end{align} 
while for antiparticle, one has
\begin{align}
v^s(|\vec{p}|=0) =\sqrt{m_\chi}
\begin{pmatrix}
 \xi^{-s} \\ 
-\xi^{-s}
\end{pmatrix} \, ,
\label{eq:v_wav}
\end{align} 
where $\xi^s$ denotes the two-component spin wavefunction. Explicitly we  have $\xi^{1/2}=(1, \, 0)^T$ and  $\xi^{-1/2} = (0, \, 1)^T$
corresponding to the DM having spin up and spin down along the $+\hat{z}$ direction. 

\section{Dressing for Monochromatic Gamma Rays}\label{sec:diphoton}

Dressing the above effective operators by closing the charged fermion $f$ into a loop, we can discuss
DM annihilation into monochromatic gamma rays: $\chi^* \chi \to \gamma \gamma , Z\gamma$ or 
$\bar \chi \chi \to \gamma \gamma , Z\gamma$. We will focus on the $\gamma\gamma$ case in this section
and present the $Z\gamma$ case in the next section. In this work, we use \texttt{FeynCalc}~\cite{Mertig:1990an,Shtabovenko:2016sxi} for analytical computation of one-loop diagrams  and \texttt{LoopTools}~\cite{Hahn:1998yk} for numerical loop integrals.

We note that the following analysis is based on the effective operators in Eqs.~\eqref{eq:effL_D1}, \eqref{eq:effL_D2},
and \eqref{eq:effL_D3a}, which involve only fermion final states. 
As a result, we only consider contributions to $\ga\ga$ and $Z\ga$ states from
a specific loop structure, {\it i.e.}, the triangle diagram in Fig.~\ref{fig:loop_ga}, which includes SM and/or new fermions only. For any UV complete theories such as supersymmetry or hidden sectors, there should exist more loop contributions than triangle diagrams. Those are, however, model-dependent and will not be discussed. Furthermore, our results will not apply to cases where the effective approach breaks down, for example, 
in the case where the mediator between the DM and SM sectors has a mass much lighter than DM. 

To illustrate the applicability of the effective theory approach,
in Appendix~\ref{sec:UV_mod} we present an exemplary UV vector portal model which can realize ${\mathcal L}^D_2$ in Eq.~\eqref{eq:effL_D2} and will give rise to same results
of $\ga\ga$ and $Z\ga$ as derived by our approach. For ${\mathcal L}^S$ and ${\mathcal L}^D_1$, 
they can be realized by Higgs portal models, while ${\mathcal L}^D_3$ may be realized by antisymmetric tensor portal models. We will not go into details for these latter models here.

\subsection{Complex Scalar Dark Matter}

For complex scalar DM annihilating into two photons via the interaction of $\mathcal L^S$, 
the calculation is virtually identical to the case of decaying DM studied by 
Elagin {\it et al}~\cite{Elagin:2017cgu}. 
In the following, we will adopt the symbols $B_0$ and $C_0$ defined in the Passarino-Veltman integrals~\cite{Passarino:1978jh},
\begin{align}
&B_0 \lee r^2_{10} , m^2_0 , m^2_1\rii = \frac{\lee 2 \pi \mu \rii^\eps}{i \pi^2}  
\int d^d k \, \prod^1_{i=0} \frac{1}{\lee k + r_i \rii^2 - m^2_i } \, , \nn \\
&C_0 \lee r^2_{10} , r^2_{12}, r^2_{20}, m^2_0 , m^2_1 , m^2_2 \rii =  \frac{\lee 2 \pi \mu \rii^\eps}{i \pi^2}  
\int d^d k \, \prod^2_{i=0} \frac{1}{\lee k + r_i \rii^2 - m^2_i }  \, ,
\end{align}
with the convention of Ref.~\cite{Mertig:1990an}, where $\eps=4-d$ and $r^2_{ij}=(r_i - r_j)^2$. In this convention, one has
\begin{align}
\lee d-4 \rii B_0 \lee 4 m_\chi^2 , m^2_f , m^2_f \rii &= -2 + \mathcal{O} \lee  \eps \rii  \, , \nn \\
C_0 \lee 4 m^2_\chi, 0, 0, m^2_f , m^2_f, m^2_f \rii & = - \frac{1}{2 \, m^2_\chi} f(x_f) \, ,
\end{align} 
where $x_f = m_f^2/m_\chi^2$ and
\begin{eqnarray}
f(x)=\left\{
\begin{array}{ll}  \displaystyle
\left( \arcsin\frac{1}{\sqrt{x}} \right)^2 & {\rm , \;\; for} \; x > 1 \; ; \\
\displaystyle -\frac{1}{4}\left[ \log\frac{1+\sqrt{1-x}}
{1-\sqrt{1-x}}-i\pi \right]^2 \hspace{0.5cm} & {\rm , \; \; for} \; x \leq 1 \; .
\end{array} \right.
\label{eq:fx}
\end{eqnarray}

\begin{figure}
\centering
\includegraphics[width=0.4\textwidth]{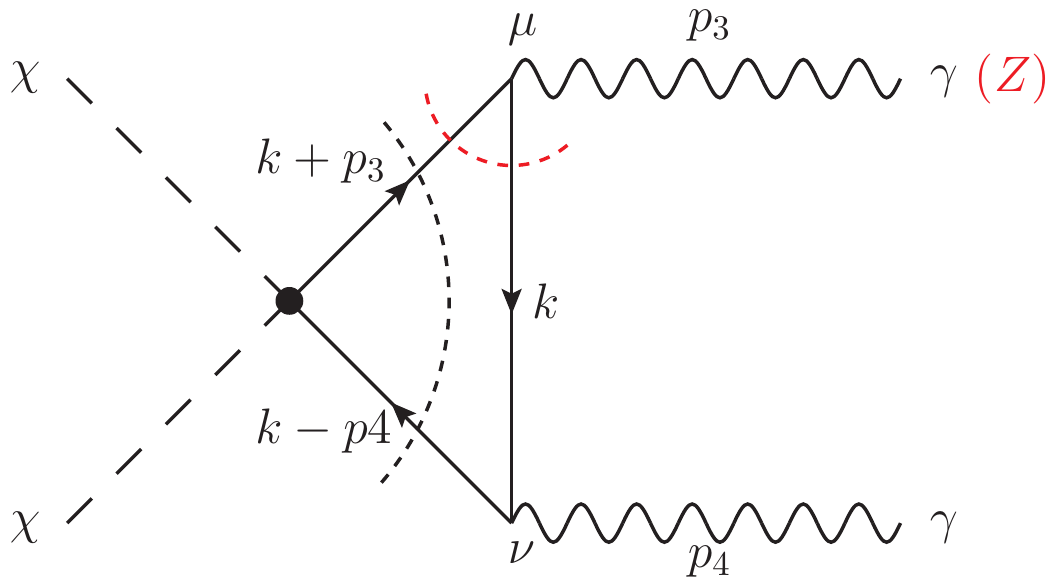} \;\; \;\; \;
\includegraphics[width=0.53\textwidth]{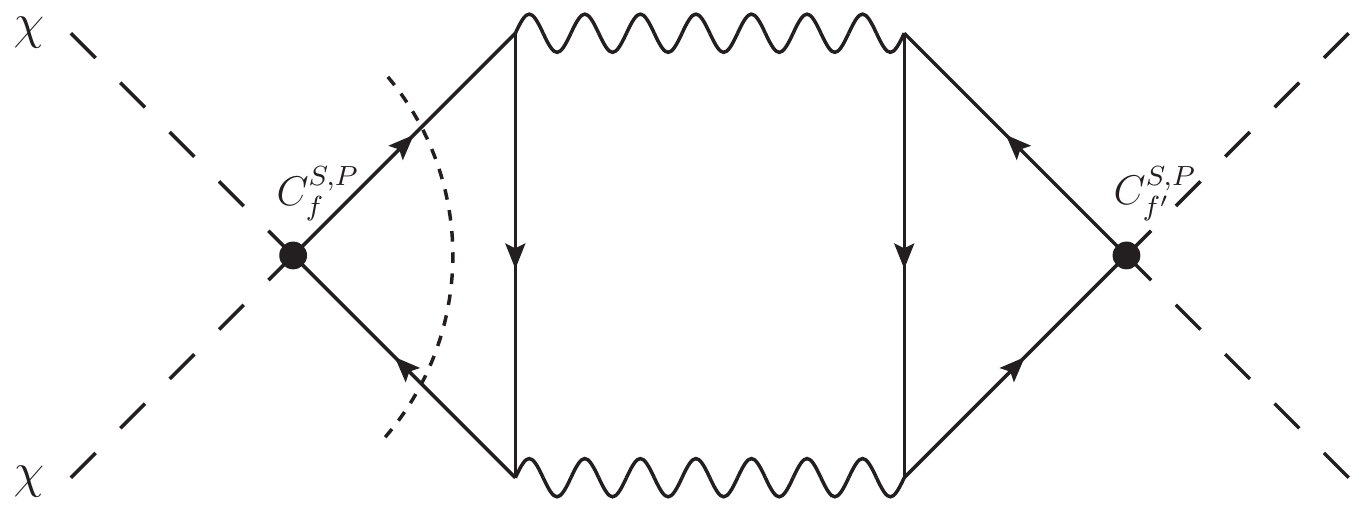}
\caption{Left: A representative loop diagram for DM annihilates into two photons, where curvy dashed lines indicate Cutkosky cuts on
internal fermions which are required for polarization asymmetry. See the main text for details.
Right: Illustration of dominant contributions from the interference of heavy-light fermions to polarization asymmetry.}
\label{fig:loop_ga}
\end{figure}

For each internal fermion species, there are two contributions:  one shown in the left panel of Fig.~\ref{fig:loop_ga}
plus the other one with the two photons being swapped, $(p_3, \, \mu) \tot (p_4, \, \nu)$.
The amplitude reads
\begin{align}
 \mathcal{M}_f =& \,
 \mathcal{M}^{\mu\nu}_f \eps^*_\mu(p_3) \eps^*_\nu(p_4) \, , \nn\\
\mathcal{M}^{\mu\nu}_f  =& - \frac{(4 \pi)^2}{\Lambda}  \, i^3  \, (-i e Q_f)^2 N_C 
 \nn\\
 \times &  \int \frac{d^d k}{ \lee 2\pi \rii^d } \text{Tr} \left(   \left( C_f^S + i C_f^P \gamma_5 \right)
\frac{  \slashed{k}  - \slashed{p}_4  +m_f }{ (k  -p_4)^2 - m^2_f} \ga^\nu 
\frac{  \slashed{k}   +m_f }{ k^2 - m^2_f} \ga^\mu
\frac{ \slashed{k} +\slashed{p}_3  +m_f }{ (k+p_3)^2 - m^2_f} \right. \nn \\
& \Biggl. \;\;\;\;\;\;\;\;\;\;\;\;\;\;\;\;\;\;\;\;\; + (p_3, \mu) \tot (p_4, \nu) 
  \Biggr) \nn\\
 =& 
- \frac{m_f}{4 \, \pi^2 \, m^2_\chi} (e Q_f)^2  N_C  \,
\Big( 
 \eta^{\mu\nu}_1 
+ 2   \, C_0\lee 4 m^2_\chi, 0, 0, m^2_f , m^2_f, m^2_f   \rii  \eta^{\mu\nu}_2
\Big)  \, ,
\end{align}
with $Q_f$ and $N_C$ being the electric charge and number of color of $f$, and 
\begin{align}
\eta^{\mu\nu}_1 =& \, C^S_f \Big(  
- 2 \, m^2_\chi g^{\mu\nu} + \lee 1 + 2 \De B_0 \rii p^\mu_3 p^\nu_4 + p^\nu_3 p^\mu_4 
\Big)
\, , \nn\\
\eta^{\mu\nu}_2 =& \, C^P_f m^2_\chi \eps^{\mu\nu p_3 p_4}   + C^S_f 
\Big(
\lee m^2_\chi - m^2_f \rii \lee 2 m^2_\chi  g^{\mu\nu}  -   p_3^\nu p_4^\mu \rii + \lee m^2_\chi + m^2_f \rii p_3^\mu p_4^\nu
\Big) \, ,
\end{align}
where
\begin{align}
\eps^{\mu\nu p_3 p_4} \equiv \eps^{\mu\nu \alpha\beta} p_{3\alpha} p_{4 \beta}  \;\; , \;\;
\De {B_0} \equiv  B_0 \lee4 m^2_\chi, m^2_f, m^2_f \rii - B_0 \lee 0, m^2_f, m^2_f\rii.
\end{align}
The polarization transversality makes vanishing contributions from the terms of $p_{3}^\mu p_{4}^\nu$ and $p_{4}^\mu p_{3}^\nu$.

Based on the definition of the photon polarization vectors in 
Eqs.~\eqref{eq:pol_vec} and \eqref{eq:pol_vec_1}, 
to leading order in $v_{\rm DM}$, the helicity amplitudes for the two photons 
in the final state are
\begin{eqnarray}
\mathcal M_f (\pm , \mp) & =  & 0 \; ,\\
\mathcal M_f (+,+) & = & \frac{( 4 \pi )^2}{ \Lambda} I_f (+, +)\; , \; \; \; 
\mathcal M_f (-,-) = \frac{( 4 \pi )^2}{ \Lambda} I_f (-, -) \; , 
\end{eqnarray} 
with
\begin{eqnarray}
\label{eq:Ipp}
\mathcal I_f (+,+) & = & \frac{ 2  N_C Q_f^2 \alpha_{\rm em} }{\pi}  m_f 
\left[
\left( \left( x_f - 1\right) f (x_f) - 1 \right) C^S_f + i f(x_f) C_f^P
\right] \; , \\
\label{eq:Imm}
\mathcal I_f (-,-) & = & \frac{ 2  N_C Q_f^2 \alpha_{\rm em} }{\pi}  m_f 
\left[
\left( \left( x_f - 1\right) f (x_f) - 1 \right) C^S_f - i f(x_f) C_f^P
\right] \; ,
\end{eqnarray}
where $\al_{\rm em}=e^2/(4 \, \pi)$, $x_f = m_f^2/m_\chi^2$ and $f(x)$ is defined in Eq.~\eqref{eq:fx}.
The result is in agreement with Ref.~\cite{Elagin:2017cgu}.
If $x_f > 1$, then $f(x_f)$ is real. This implies $\mathcal I_f(-,-) = ( \mathcal I_f(+,+) )^*$, and thus there will be no net circular
polarization even though CP is violated by complex couplings. 
On the other hand, if $x_f \leq 1$, then $f(x_f)$ is complex and in general
$\vert \mathcal I_f (+,+) \vert^2 \neq \vert \mathcal I_f (-,-) \vert^2$. It corresponds to the internal fermions being on-shell
marked by the black dashed line~(Cutkosky cut) in the left panel of Fig.~\ref{fig:loop_ga}.
As a result, a net circular polarization can be produced in this case.
Note that
there are in fact three possibilities of having complex loop integrals which are correlated with Cutkosky cuts on any two of the three internal fermions.
As we shall see below, when one of the outgoing photons is replaced by the massive $Z$ boson, there exists more 
nontrivial region of the parameter space that can induce polarization symmetry, as indicated by the red
dashed Cutkosky cut on the internal fermions connected to $Z$.

In reality, we need to sum over all possible charged fermions running inside the loop.
\begin{equation}
\vert \mathcal M \vert^2 = \biggl\vert \sum_f \mathcal M_f (+, +) \biggr\vert^2 + \biggl \vert \sum_f \mathcal M_f (-, -) \biggr\vert^2 \; .
\end{equation}
Therefore, there are mixing terms between contributions from different fermions leading to complicated expressions of the asymmetry between
the $(+,+)$ and $(-,-)$ polarizations.
Although directly from Eqs.~\eqref{eq:Ipp} and \eqref{eq:Imm} it is straightforward to infer the total asymmetry including all contributions,
for illustrative purposes we show only the asymmetry due to a single fermion field $f$ in the loop:
\begin{align}
\label{eq:asym_th}
 \left( \biggl\vert  \mathcal M_f (+, +) \biggr\vert^2 - \biggl \vert  \mathcal M_f (-, -) \biggr\vert^2  \right) =
  8 \lee \frac{  N_C Q_f^2 \alpha_{\rm em} }{\pi}  m_f \rii^2 
 C^2_f \, \vert f(x_f) \vert
 \sin2\th_1 \sin\th_2 \; ,
\end{align}
where
\begin{align}
C_f= \sqrt{ {C^S_f}^2 + {C^P_f}^2 }
\;\; , \;\;
C^P_f =C_f \sin\th_1
\;\; \text{and} \;\;
f(x_f) = \vert f(x_f)\vert \lee \cos\th_2 + i \sin\th_2 \rii \, .
\end{align}
It is clear that the asymmetry requires two complex phases associated with the coupling constants and  
the loop integrals, respectively. In addition, both $C^S_f$ and $C^P_f$ have to be nonzero; otherwise, one can always rotate away the phase
$\th_1$ by field redefinition.

Due to the fact that the amplitude is proportional to $m_f$,
in case DM is heavier than all of possible internal fermions and the effective couplings are roughly the same order
for the fermions, dominant contributions to polarization asymmetry will come from the heaviest fermions.
On the other hand, when the DM mass is among the fermion ones, 
the main contributions arise from the heavy-light interference where the heavy internal fermion provides a complex coupling constant while the light-fermion
loop supplies a complex loop integral as demonstrated in the right panel of Fig.~\ref{fig:loop_ga}.

\subsection{Dirac Dark Matter}

Here we consider effective operators $\mathcal L^D_{1,2,3}$ for Dirac DM. 
Consider $\chi (k_1, s_1) \bar \chi (k_2, s_2) \to \gamma (k_3, h_3) \gamma (k_4, h_4)$,
where $(k_1, s_1)$ and $(k_2 , s_2)$ are the momenta/helicities of the DM $\chi$ and $\bar \chi$, 
and $(k_3, h_3)$ and $(k_4 , h_4)$ are the momenta/helicities of the two final state photons.
Denote the helicity amplitude for the process as $\mathcal M(s_1,s_2;h_3,h_4)$. 
Summing over the initial spins of the DM, we have
\begin{equation}
\vert \mathcal M (h_3 , h_4 ) \vert^2 = \sum_{s_1, s_2} \vert \mathcal M(s_1,s_2  ;  h_3,h_4) \vert^2 \; .
\end{equation}

\subsubsection{$\mathcal L^D_1$}

The calculation for this case is very similar to $\mathcal L^S$. 
\begin{equation}
 \vert \mathcal M (h_3 , h_4 ) \vert^2 =
 \frac{( 4 \pi )^4 }{\Lambda^4} 
\left[ \left( k_1 \cdot k_2 - m_\chi^2 \right) ( C^S_\chi )^2 + \left( k_1 \cdot k_2 + m_\chi^2 \right) ( C^P_\chi )^2 \right]
\biggl\vert \sum_f \mathcal I_f \left( h_3 , h_4 \right) \biggr\vert^2 \, .
\end{equation} 
$\mathcal I_f(h_3, h_4)$ is the helicity amplitude of the two photon final state. Again to leading order in $v_{\rm DM}$, 
the only non-vanishing components are 
$\mathcal I_f(+, +)$ and $\mathcal I_f(-, -)$ given by \eqref{eq:Ipp} and \eqref{eq:Imm} respectively.
For non-relativistic DM, $k_1 = k_2 \approx (m_\chi, \vec 0 )$ and $k_1 \cdot k_2 = m_\chi^2$. We then have the simpler result
\begin{equation}
 \vert \mathcal M (h_3 , h_4 ) \vert^2  \approx 
  \frac{( 4 \pi )^4 }{\Lambda^4}  2 \, m_\chi^2 ( C^P_\chi )^2 
\biggl\vert \sum_f \mathcal I_f \left( h_3 , h_4 \right) \biggr\vert^2 \; .
\end{equation} 
Note that $C^S_\chi$ has dropped out in the non-relativistic limit because of velocity suppression.
Again since in general we have $\vert \sum_f \mathcal I_f(+,+) \vert^2 \neq \vert \sum_f \mathcal I_f(-,-) \vert^2$, net circular 
polarization will be produced from $\chi \bar \chi \to \gamma \gamma$ via $\mathcal L^D_1$.
The circular polarization asymmetry from a single fermion species is proportional to 
\begin{align}
 \frac{1}{4} \lee \biggl\vert  \mathcal M_f (+, +) \biggr\vert^2 - \biggl \vert  \mathcal M_f (-, -) \biggr\vert^2 \rii =
 16 \lee \frac{  N_C Q_f^2 \alpha_{\rm em} }{\pi}  m_f  m_\chi  \rii^2 
 {C^P_\chi}^2 \, C^2_f  \, \vert f(x_f) \vert
 \sin2\th_1 \sin\th_2 \; ,
\end{align}
which is identical to Eq.~\eqref{eq:asym_th} up to a factor of $2 \, m^2_\chi {C^P_\chi}^2$.

\subsubsection{$\mathcal L^D_2$}

Although it is not expected that $\mathcal L^D_2$ will give rise to net circular polarization as all the couplings are real, we however check this statement by 
explicit calculation.
For $\chi(p_1) \, \bar{\chi}(p_2) \to \ga(\eps^\mu(p_3) ) \, \ga(\eps^\nu(p_4))$, the DM side has the amplitude 
\begin{align}
\mathcal{M} _\sig({\text{DM}}) =   \frac{\left(4 \pi\right)^2 }{\Lambda^2}\sum_{s,s'}
\overline{v^{s'}(p_2)} \ga_\sig \lee C^L_\chi P_L + C^R_\chi P_R \rii u(p_1)^s \, ,
\end{align}
while the SM side has
\begin{align}
 \mathcal{M}_f^\sigma({\text{SM}}) =& \,
 \mathcal{M}_f^{\sigma\mu\nu} \eps^*_\mu(p_3) \eps^*_\nu(p_4) \, , \nn\\
\mathcal{M}_f^{\sigma\mu\nu}  = & \, (-1) \, i^3  \, (-i e Q_f)^2 N_C \frac{\lee C^R_f - C^L_f \rii}{2}  \, 
 \nn\\
& \times \int \frac{d^d k}{ \lee 2\pi \rii^d } \text{Tr} \left( \ga^\sig \ga^5
\frac{  \slashed{k}  - \slashed{p}_4  +m_f }{ (k  -p_4)^2 - m^2_f} \ga^\nu 
\frac{  \slashed{k}   +m_f }{ k^2 - m^2_f} \ga^\mu
\frac{ \slashed{k} +\slashed{p}_3  +m_f }{ k^2 - m^2_f} \right. \nn \\
& \Biggl. \;\;\;\;\;\;\;\;\;\;\;\;\;\;\;\;\;\;\;\;\;\;\;\;\;\;\; + (p_3, \mu) \tot (p_4, \nu) 
  \Biggr) \, ,\nn\\
 =& \, (-i) \, (e Q_f)^2  N_C \frac{\lee C^R_f - C^L_f \rii}{2}  \nonumber \\
& \times \left[
 \la^{\sigma\mu\nu}_1 \De B_0 
+ \la^{\sigma\mu\nu}_2  \lee 1+ 2 \,  m^2_f  \, C_0\lee 4 m^2_\chi, 0, 0, m^2_f , m^2_f, m^2_f \rii \rii
\right] \, ,
\end{align}
with
\begin{align}
\la^{\sigma\mu\nu}_1=& \, \frac{1}{8 \pi^2 \, m^2_\chi } 
\lee 
\eps^{\sigma \nu p_3 p_4 }  p^\mu_3 - \eps^{\sigma \mu p_3 p_4 }  p^\nu_4
\rii  \, ,
  \nn\\
\la^{\sigma\mu\nu}_2=& \, \frac{1}{8 \pi^2 \, m^2_\chi }
\lee 
\eps^{\sigma \mu p_3 p_4 }  p^\nu_3 - \eps^{\sigma \nu p_3 p_4 }  p^\mu_4
+ 2 m^2_\chi \lee  \eps^{\sigma \mu \nu p_3}  - \eps^{\sigma \mu \nu p_4 }  \rii
\rii    \, ,
\end{align}
where $\eps^{\sig \mu \nu p_{(3,4)}}= \eps^{\sig \mu\nu \kappa}  p_{(3,4)\kappa}$. 
Again, the $\la_1^{\sigma\mu\nu}$ term will not contribute as transversality of the photon polarization vectors dictates
$p^\mu_3 \eps_\mu(p_3)=p^\nu_4 \eps_\nu(p_4)=0$,
which is not the case if the photon is replaced by the $Z$ boson that has a longitudinal component.
Note that the term proportional $C^R_f + C^L_f $~($\ga^\sigma$ term)
is vanishing due to Furry's theorem from charge conjugation ($C$) invariance in QED.
In addition, our results are consistent with Ref.~\cite{Rosenberg:1962pp}, which
demonstrates that the amplitude $ \mathcal{M}_f^{\sigma\mu\nu}$
can in general be decomposed into six terms. The corresponding coefficients are correlated as a result of the invariance under the two photon exchange,
$(\mu, p_3) \tot (\nu, p_4)$, and the Ward-Takahashi identity: $ \mathcal{M}_f^{\sigma\mu\nu} p_{3\mu}= \mathcal{M}_f^{\sigma\mu\nu} p_{4\nu}=0$.
See, for instance, Stephen L. Adler's lecture on ``Perturbation Theory Anomalies'' in~\cite{Deser:1970spa} for a pedagogical review.

Furthermore, one can reproduce computation of the well-known axial current anomaly by contracting the amplitude with the total momentum
in the limit of $m_f \to 0$:
\begin{align}
i (p_3+p_4)_\sigma \mathcal{M}_f^\sigma({\text{SM}}) = - (e Q_f)^2  N_C \frac{\lee C^R_f - C^L_f \rii}{2} 
\frac{\eps^{\eps^*_3 \eps^*_4 p_3 p_4}}{2 \pi^2} \, .
\end{align} 
By combining the DM and SM amplitudes and with the help of Eqs.~\eqref{eq:pol_vec} to \eqref{eq:v_wav},
one has for the amplitude squared
\begin{equation}
\overline{\vert \mathcal M \vert^2} = \frac{1}{4} \left( \biggl\vert \sum_f \mathcal M_f (+, +) \biggr\vert^2 + \biggl \vert \sum_f \mathcal M_f (-, -) \biggr\vert^2 \right) \; ,
\end{equation}
where to leading order in $v_{\rm DM}$,
\begin{align}
\label{eq:gagaA}
\mathcal{M}_f (\pm, \mp)&= 0  \, , \\
 \mathcal{M}_f (\pm, \pm)&=  
 \frac{ \lee 4 \pi \rii^2 }{\Lambda^2}
 \frac{ \sqrt{2} \, N_C \, Q_f^2 \, \alpha_{\rm em} m^2_\chi }{\pi} \lee C^R_\chi - C^L_\chi \rii \lee C^R_f - C^L_f \rii 
 \lee 1 -  x_f f(x_f)\rii  \, . 
\end{align}
Since  $\vert \sum_f M_f (+, +) \vert^2=\vert \sum_f M_f (-, -) \vert^2$, there is  no net circular polarization as expected from general argument
of the lack of complex couplings in $\mathcal L^D_2$.

\subsubsection{$\mathcal L^D_3$}

For $\mathcal L^D_3$, the calculation is straightforward but tedious.
The DM side has the amplitude 
\begin{align}
\mathcal{M} _{\al\beta}({\text{DM}}) =   \frac{\left(4 \pi\right)^2 }{\Lambda^2}\sum_{s,s'}
\overline{v^{s'}(p_2)} 
\sigma_{\al \beta}  \left( \widetilde C_\chi^{L} P_L + \widetilde C_\chi^{R} P_R \right)
 u(p_1)^s \, ,
\end{align}
while the SM side has 
\begin{align}
 \mathcal{M}_f^{\al\beta} ({\text{SM}}) =& \, \mathcal{M}_f^{\alpha\beta\mu\nu} \eps^*_\mu(p_3) \eps^*_\nu(p_4) \, , \nn\\
 \mathcal{M}_f^{\alpha \beta\mu\nu} =& \, (-1) \, i^3  \, (-i e Q_f)^2 N_C  \frac{\lee \widetilde C^R_f - \widetilde C^L_f \rii}{2} 
 \nn\\
\times &  \int \frac{d^d k}{\lee 2\pi \rii ^d} \text{Tr} \left( \sigma^{\al\beta} \ga^5
\frac{   \slashed{k}  - \slashed{p}_4  +m_f }{ (k  -p_4)^2 - m^2_f} \ga^\nu 
\frac{  \slashed{k}  +m_f }{ k^2 - m^2_f} \ga^\mu
\frac{ \slashed{k} + \slashed{p}_3 +m_f }{ (k + p_3)^2 - m^2_f} \right. \nn \\
 & \Biggl. \;\;\;\;\;\;\;\;\;\;\;\;\;\;\;\;\;\;\;\;\; + (p_3, \mu) \tot (p_4, \nu) 
  \Biggr) \, ,\nn\\
  =& \, (-i) (e Q_f)^2  N_C \frac{\lee \widetilde C^R_f - \widetilde C^L_f \rii}{2}  \,
  \ka_1^{\alpha \beta\mu\nu} C_0 \lee 4 m^2_\chi, 0, 0, m^2_f , m^2_f, m^2_f \rii \, ,
\end{align}
with
\begin{align}
\ka_1^{\alpha \beta\mu\nu} =  \frac{m_f}{4 \pi^2  }
\lee
g^{\al\mu} \eps^{\beta\nu p_3 p_4} - g^{\al\nu} \eps^{\beta\mu p_3 p_4}
+ p_3^\al \eps^{\beta\mu\nu p_4} - p_4^\al \eps^{\beta\mu\nu p_3}
- \lee \al \tot \beta \rii  
  \rii \, .
\end{align}
Based on Eqs.~\eqref{eq:pol_vec}, \eqref{eq:u_wav} and \eqref{eq:v_wav}, 
it is straightforward to show that $\ka_1^{\alpha \beta\mu\nu}$ is zero after contracting with $\eps^*_\mu(p_3) \eps^*_\nu(p_4)$
and $\sigma_{\al \beta}$ on indices $\al$, $\beta$,
$\mu$ and $\nu$. 

A simple way to understand the vanishing amplitude is to notice
that the initial state corresponding to either the magnetic or electric dipole moment is odd under $C$,
whereas each of the two photons in the final state is also odd under $C$. It leads to a vanishing amplitude 
according to the Furry's theorem.

Another way to understand this is to see if one can write down effective operators to describe the amplitude. Due to gauge invariance one needs to 
involve two EM field strength. 
Possible effective operators are 
\begin{eqnarray}
&&  \bar \chi \sigma_{\mu \nu} \left( \widetilde C_\chi^{L} P_L + \widetilde C_\chi^{R} P_R \right) \chi  \,
F^{\mu \alpha} F^{\nu}_\alpha  \, , \\ 
&&   \bar \chi \sigma_{\mu \nu} \left( \widetilde C_\chi^{L} P_L + \widetilde C_\chi^{R} P_R \right) \chi  \,
 \epsilon^{\mu\nu\rho\sigma} F_{\rho\alpha} F_\sigma^\alpha \, ,\\ 
&&   \bar \chi \sigma_{\mu \nu} \left( \widetilde C_\chi^{L} P_L + \widetilde C_\chi^{R} P_R \right) \chi  \,
F^{\mu \alpha} \widetilde F^{\nu}_\alpha \, .
\end{eqnarray}  
The first two operators vanish identically.
The third operator is non-zero, but as demonstrated by explicit calculation above, its coefficient is zero.

\section{$Z\ga$ Final State}~\label{sec:Z_gamma}

Here we collect the results for the $Z\ga$ final state, where we sum over three $Z$ polarizations:
right-handed~($+$), left-handed~($-$) and longitudinal~($L$) polarizations. Note that for the following results,
we have explicitly checked that 
the Goldstone boson equivalence theorem~(for $Z(p_3)$) 
and the Ward-Takahashi identity~(for $\ga(p_4)$) 
hold respectively. That is,
$\mathcal{M^{\al\cdots \mu\nu}} p_{3\mu}=m_Z \mathcal{M}^{\al\cdots\nu}$
and
$\mathcal{M^{\al\cdots \mu\nu}} p_{4\nu}=0$, where $m_Z$ is the $Z$ mass.

\subsection{Complex Scalar Dark Matter}

Due to conservation of angular momentum,  only the $(+,+)$ and $(-,-)$ helicity configurations 
with the first~(second) entry refers to the $Z~(\ga)$ polarization
have nonzero amplitudes.
\begin{equation}
\vert \mathcal M \vert^2 = 
\biggl\vert \sum_f \mathcal M_f (+, +) \biggr\vert^2 
+ \biggl \vert \sum_f \mathcal M_f (-, -) \biggr\vert^2 
 \; ,
\end{equation}
where to leading order in $v_{\rm DM}$, $\mathcal M_f (\pm, \mp) = \mathcal M_f(L, \pm) = 0$ and
\begin{align}
 \mathcal M_f (+,+)
&= \frac{2 y_s}{\pi} \frac{ \lee 4 \pi\rii^2 }{\La}
\frac{ 4- x_Z}{4}
m_f \left[
- \frac{\rho_1}{(4 -x_Z)^2} {C^S_f}  + i \rho_2 {C^P_f}
\right] \; , 
\nn\\
\mathcal  \mathcal M_f (-,-) 
&=  \frac{2 y_s}{\pi} \frac{ \lee 4 \pi\rii^2 }{\La}
\frac{ 4- x_Z}{4}
m_f \left[
- \frac{\rho_1}{(4 -x_Z)^2} {C^S_f}  - i \rho_2 {C^P_f}
\right]
 \; , 
\end{align}
with
\begin{align}
\rho_1=& \, 16 +  g(x_f,x_Z)  \lee 4 - x_Z \rii  \lee 4 - 4 \, x_f - x_Z \rii + 4 \, x_Z \De B_1 \; ,
 \nn\\
\rho_2=&  \, g(x_f,x_Z) \; ,
\label{eq:rho_12}
\end{align}
where $x_Z = m_Z^2/m_\chi^2$, $y_s=  N_C \, Q^2_f \, z_V \, \alpha_{\rm em}$ 
(with the vector coupling $z_V$ defined in Eq.~{\eqref{zV}} below) and
\begin{align}
g(x_f,x_Z)  \equiv & \, - 2 \, m^2_\chi  C_0 \lee 4 m^2_\chi,  m^2_Z, 0, m^2_f , m^2_f, m^2_f \rii \, , \nn\\
\De B_1 \equiv &  \, B_0 \lee4 m^2_\chi, m^2_f, m^2_f \rii - B_0 \lee m^2_Z, m^2_f, m^2_f \rii  - 1 \, .
\label{eq:fxz}
\end{align}
Note that $g(x_f,x_Z)$ and $\De B_1$ can be complex if the internal fermions are on-shell, where complex
$B_0 \lee4 m^2_\chi, m^2_f, m^2_f \rii$ and $B_0 \lee m^2_Z, m^2_f, m^2_f \rii$ correspond to two Cutkosky cuts marked by the black and red dashed lines
respectively in the left-panel of Fig.~\ref{fig:loop_ga}.

The symbol $z_V$ corresponds to the fermion vector coupling\footnote{The $Z$ axial current does not contribute due to the Furry's theorem as the axial current is even under $C$.} to $Z$ normalized to the corresponding
electric charge.
For instance, with the internal electron one has
\begin{align}
\label{zV}
z_V= \frac{1}{e Q_e}\frac{g}{\cos\th_W} \lee \frac{T^3 - 2\sin^2\th_W Q_e}{ 2} \rii \, ,
\end{align}
where $g$ is the SM $SU(2)_L$ gauge coupling, $e$ is the electric coupling, $\th_W$ is the Weinberg angle,
$Q_e=-1$ and $T^3 = -1/2$. In the following, we will also use the symbol $z_A$ for the axial current, {\it e.g.},
\begin{align}
\label{zA}
z_A= \frac{1}{e Q_e}\frac{g}{\cos\th_W} \lee - \frac{T^3}{ 2} \rii \, ,
\end{align}
for the electron.
Note that one can reproduce Eqs.~\eqref{eq:Ipp} and \eqref{eq:Imm} above by setting $x_Z=0$, $z_V=1$,
and $g(x_f,x_Z) \to f(x_f)$.
The circular polarization asymmetry from contributions of a fermion $f$ is proportional to 
\begin{align}
\label{asyScalarZga}
& \biggl\vert  \mathcal M_f (+, +) \biggr\vert^2 - \biggl \vert  \mathcal M_f (-, -) \biggr\vert^2  \nn\\
 &=
  \frac{ \lee 4 \pi\rii^4 }{\La^4} \frac{ 2 \,  y^2_s \, x_f}{ \pi^2 }
 C^2_f \, \vert g(x_f,x_Z) \vert
 \sin2\th_1  \lee 
 4\sin\th'_2 + x_Z \vert \De B_1 \vert \sin\lee \th'_2 -\th_3 \rii  
 \rii  \; ,
\end{align}
where
\begin{align}
g(x_f,x_Z) = \vert g(x_f,x_Z)\vert \lee \cos\th'_2 + i \sin\th'_2 \rii
\;\; \text{and} \;\;
\De B_1 = \vert \De B_1 \vert \lee \cos\th_3 + i \sin\th_3 \rii
 \, .
\end{align}

\subsection{Dirac Dark Matter}

\subsubsection{$\mathcal L^D_1$}

As mentioned above in the diphoton case, at amplitude-squared level 
the result is similar to that of the scalar DM case, except for an additional factor from
the trace of  DM spinor wavefunctions.
In the limit of zero DM velocity, the asymmetry is simply given by Eq.~\eqref{asyScalarZga} multiplied by 
$2 m^2_\chi (C^P_\chi)^2$.

\subsubsection{$\mathcal L^D_2$}

One has for the amplitude squared

\begin{equation}
\overline{\vert \mathcal M \vert^2} = \frac{1}{4} \lee
\biggl\vert \sum_f \mathcal M_f (+, +) \biggr\vert^2 + \biggl\vert \sum_f \mathcal M_f (L, +) \biggr\vert^2 +
\biggl\vert \sum_f \mathcal M_f (-, -) \biggr\vert^2 + \biggl\vert \sum_f \mathcal M_f (L, -) \biggr\vert^2 
\rii
 \, ,
\end{equation}
where $\vert \mathcal M_f (+, +)\vert^2 = \vert \mathcal M_f (-, -)\vert^2$ and $\vert \mathcal M_f (L, +)\vert^2 = \vert \mathcal M_f (L, -)\vert^2$.
For contributions from a  fermion $f$, we obtain, 
to leading order in $v_{\rm DM}$, $\mathcal M_f (\pm, \mp) = 0$ and
\begin{align}
\vert \mathcal M_f (+, +)\vert^2 =& \, y^2_V \frac{\lee 4 \pi\rii^4}{\La^4} \frac{1}{8 \, \pi^2}
\lee
\frac{ \vert \ka^f_1 \vert^2}{\lee 4 - x_Z\rii^2 } \lee C^R_\chi+C^L_\chi \rii^2   + \vert \ka^f_2 \vert^2 \lee C^R_\chi - C^L_\chi \rii^2
 \rii \, ,
 \nn\\
 \vert \mathcal M_f (L, +)\vert^2 =& \, y^2_V \frac{\lee 4 \pi\rii^4}{\La^4} \frac{1}{2 \, \pi^2}
\frac{  \vert \ka^f_1 \vert^2 }{ x_Z \lee 4 - x_Z \rii^2} \lee C^R_\chi+C^L_\chi \rii^2
\, ,
\end{align}
with $y_V=  N_C \, Q^2_f \, \alpha_{\rm em} m^2_\chi$.
The coefficients $\ka^f_1$ and $\ka^f_2$ are 
\begin{align}
\ka^f_1=&  
\lee C^R_f+C^L_f  \rii z_A \lee - 4 \, g(x_f,x_Z) \, x_f \lee 4 -x_Z \rii + x_Z \lee 8 - x_Z + 4 \De B_1\rii \rii   \nn \\
& + \lee C^R_f - C^L_f  \rii x_Z z_V \lee 8 - g(x_f,x_Z) \, x_f \lee 4 -x_Z \rii - x_Z  + 4 \De B_1 \rii     \; ,
\nn \\
\ka^f_2=& \lee 4 - x_Z \rii 
\lee  
\lee C^R_f+C^L_f  \rii z_A +  \lee C^R_f - C^L_f  \rii z_V \lee 1 - g(x_f,x_Z) x_f  \rii
\rii \, .
\end{align}

It is straightforward to generalize to cases with more than one fermion by the replacement
\begin{align*} 
Q^2_f \ka^f_{i} \to \sum_{f} Q^2_{f} \ka^{f}_{i} \; ,
\end{align*}
for $i=(1,2,3)$.
Note again that by setting $z_A=0$, $x_Z=0~(m_Z=0)$ and $z_V=1$ which implies $\ka^f_1=0$ and hence the longitudinal component drops,
Eq.~\eqref{eq:gagaA} is reproduced. Due to Furry's theorem, 
the contribution from the SM fermion vector current~(axial current) is nonzero only in the presence of the
the $Z$ axial current~(vector current). 
Nevertheless just like the diphoton case 
there is no asymmetry in this $Z$-photon case as well 
due to the lack of complex couplings for $CP$ violation in ${\mathcal L}^D_2$.

\subsubsection{$\mathcal L^D_3$}

Similarly, one has for the amplitude squared

\begin{equation}
\overline{\vert \mathcal M \vert^2} = \frac{1}{4} \lee
\biggl\vert \sum_f \mathcal M_f (+, +) \biggr\vert^2 + \biggl\vert \sum_f \mathcal M_f (L, +) \biggr\vert^2 +
\biggl\vert \sum_f \mathcal M_f (-, -) \biggr\vert^2 + \biggl\vert \sum_f \mathcal M_f (L, -) \biggr\vert^2 
\rii
 \, .
\end{equation}
For contributions from a fermion $f$, one has to leading order in $v_{\rm DM}$,
\begin{align}
\vert \mathcal M_f (+, +)\vert^2 =& \, y^2_A \frac{\lee 4 \pi\rii^4}{\La^4} \frac{ 16 \, x_f }{ \lee 4 - x_Z\rii^2 \pi^2}
\left| \sqrt{2} \widetilde{C}^*_{\chi f}  \la_1  + \frac{1}{\sqrt 2}\widetilde{C}_{\chi f} \la_2 \right|^2 \, 
 \nn\\
 =& \, y^2_A \frac{\lee 4 \pi\rii^4}{\La^4} \frac{ 16 \, x_f }{ \lee 4 - x_Z\rii^2 \pi^2}
\lee
\lee \widetilde{C}^2_{\chi f}  \la^*_1 \la_2 +  {\widetilde{C}^{*2}_{\chi f}} \la_1 \la^*_2 \rii 
+ \frac{\widetilde{C}_{\chi f}  {\widetilde{C}^{*}_{\chi f}} }{2} \lee 4\vert \la_1\vert^2 + \vert \la_2\vert^2 \rii  
 \rii \, ,
 \nn\\
 \vert \mathcal M_f (-, -)\vert^2 =& \, y^2_A \frac{\lee 4 \pi\rii^4}{\La^4} \frac{ 16 \, x_f }{ \lee 4 - x_Z\rii^2 \pi^2}
\left| \sqrt{2} \widetilde{C}_{\chi f}  \la_1  + \frac{1}{\sqrt 2}\widetilde{C}^*_{\chi f} \la_2 \right|^2 \,
 \nn\\
=& \, y^2_A \frac{\lee 4 \pi\rii^4}{\La^4} \frac{ 16 \, x_f }{ \lee 4 - x_Z\rii^2 \pi^2}
\lee
\lee \widetilde{C}^2_{\chi f}  \la_1 \la^*_2 +  {\widetilde{C}^{*2}_{\chi f}} \la^*_1 \la_2 \rii 
+ \frac{\widetilde{C}_{\chi f}  {\widetilde{C}^{*}_{\chi f}} }{2} \lee 4\vert \la_1\vert^2 + \vert \la_2\vert^2 \rii  
 \rii \, ,
 \nn\\
\vert \mathcal M_f (L, +)\vert^2 =& \, y^2_A \frac{\lee 4 \pi\rii^4}{\La^4} \frac{ 32 \, x_f}{ \lee 4 - x_Z\rii^2 \pi^2}
\left| 2 \widetilde{C}^*_{\chi f}  \la_1  + \frac{1}{ 2}\widetilde{C}_{\chi f} \la_3 \right|^2
 \,  \nn \\
  =& \, y^2_A \frac{\lee 4 \pi\rii^4}{\La^4} \frac{ 32 \, x_f}{ \lee 4 - x_Z\rii^2 \pi^2}
\lee
\lee \widetilde{C}^2_{\chi f}  \la^*_1 \la_3 +  {\widetilde{C}^{*2}_{\chi f}}  \la_1 \la^*_3 \rii 
+ \frac{\widetilde{C}_{\chi f}  {\widetilde{C}^{*}_{\chi f}} }{4} \lee 16 \vert \la_1 \vert^2 + \vert \la_3 \vert^2 \rii  
 \rii \, ,
 \nn \\
 \vert \mathcal M_f (L, -)\vert^2 =& \, y^2_A \frac{\lee 4 \pi\rii^4}{\La^4} \frac{ 32 \, x_f}{ \lee 4 - x_Z\rii^2 \pi^2}
\left| 2 \widetilde{C}_{\chi f}  \la_1  + \frac{1}{ 2}\widetilde{C}^*_{\chi f} \la_3 \right|^2
 \,  \nn \\
 =& \, y^2_A \frac{\lee 4 \pi\rii^4}{\La^4} \frac{ 32 \, x_f}{ \lee 4 - x_Z\rii^2 \pi^2}
\lee
\lee \widetilde{C}^2_{\chi f}  \la_1 \la^*_3 +  {\widetilde{C}^{*2}_{\chi f}}  \la^*_1 \la_3 \rii 
+ \frac{\widetilde{C}_{\chi f}  {\widetilde{C}^{*}_{\chi f}} }{4} \lee 16 \vert \la_1 \vert^2 + \vert \la_3 \vert^2 \rii  
 \rii \, , \nn \\
 \mathcal M_f (\pm,\mp) = & \, 0 \, , 
\end{align}
with $y_A=  N_C \, z_A \, Q^2_f \, \alpha_{\rm em} m^2_\chi$ and
\begin{align}
\la_1 =& \, 4 - x_f \, g(x_f,x_Z) \lee 4 - x_Z \rii + x_Z \De B_1 \, , \nn \\
\la_2 =& \, g(x_f,x_Z) \lee 4 - x_Z \rii \lee -4 - 2 x_f + x_Z \rii  - 2 \, x_Z \lee 2 + \De B_1 \rii  + 8 \lee 3 + 2 \De B_1 \rii  \, ,  \\
\la_3 =& \, -16 \lee  1 + \De B_0 + \De B_1 \rii \nn \\
& \;\;\;\;\;\;\;\;\; + x_Z \left[  12 - x_f g(x_f,x_Z) \lee 4 - x_Z  \rii 
 + 8 \De B_0 - x_Z \lee 1 + \De B_0 \rii + 8\De B_1  \right] \, . 
\end{align}
Similarly, it is straightforward to generalize to cases of multiple fermions by factoring in $z_A$ and $m_f$~(terms depending on fermion properties) and summing up contributions
within $\vert \;\; \vert^2$, {\it i.e.}, $\vert {\rm Stuff}_f \vert^2 \to \vert \sum_f {\rm Stuff}_f \vert^2$.
It is clear by setting $z_A=0$, the amplitude vanishes as in the two photon final state. However, with the $Z\ga$ final state, the polarization asymmetry can be generated if $\widetilde{C}$s are complex and the internal particles are on-shell.

For illustration, the circular polarization asymmetry resulting from a single fermion contribution is proportional to 
\begin{align}
&  \frac{1}{4} \lee  \biggl\vert  \mathcal M_f (+, +) \biggr\vert^2
+ \biggl\vert  \mathcal M_f (L, +) \biggr\vert^2
 - \biggl \vert  \mathcal M_f (-, -) \biggr\vert^2 
  - \biggl \vert  \mathcal M_f (L, -) \biggr\vert^2 \rii 
 \nn\\
 & \;\;\;\;\;\; \;\; =
\frac{ \lee 4 \pi\rii^4 }{\La^4} \frac{ 16 \,  y^2_A \, x_f }{ \pi^2  \, x_Z}
 \vert \widetilde{C}_{\chi f} \vert^2 \sin\lee 2 \,\th_{\chi f}   \rii \lee 2 t_1 + t_2 + 2 t_3 \rii \, ,
 \end{align}
where
\begin{align}
\widetilde{C}_{\chi f}=& \, \vert \widetilde{C}_{\chi f} \vert \lee \cos\th_{\chi f} + i \sin\th_{\chi f} \rii \, ,\nn\\
t_1=& \, \vert g(x_f,x_Z) \vert \sin\th'_2 \lee 2 x_Z + x_f \lee 4 - 3 x_Z \rii \rii \, , \nn\\
t_2=& \, \vert g(x_f,x_Z) \vert \, \vert \De B_1 \vert \sin(\th'_2 - \th_3) \lee x^2_Z  + x_f \lee 8 -6 x_Z \rii \rii  \, ,\nn\\
t_3=& \, \vert \De B_1 \vert  \sin\th_3 \lee 4 - 3 x_Z \rii +   
\vert \De B_0 \vert 
\Big(
x_f \, \vert  g(x_f,x_Z) \vert \sin(\th'_2 - \th_4) \lee 4 - x_Z \rii
\nn\\
& - x_Z \vert \De B_1 \vert \sin(\th_3 - \th_4)  + 4 \sin\th_4
\Big) \, ,
\end{align}
and 
$\De B_0 = \vert \De B_0 \vert \lee \cos\th_4 + i \sin\th_4 \rii$.

Here we summarize our theoretical calculation.  In the diphoton final state, only two effective operators $\mathcal L^S$ and $\mathcal L^D_1$
can give rise to circular polarization asymmetry, whereas in the $Z$-photon final state, besides $\mathcal L^S$ and $\mathcal L^D_1$, 
$\mathcal L^D_3$ can also generate the asymmetry.
It is necessary in each non-vanishing case to have couplings with $P$ and $CP$ violation and some internal particles have to 
go on-shell (Cutkosky cut) to generate the asymmetry.

\section{Numerical Results}\label{sec:results}

\begin{figure}
\centering
\includegraphics[width=0.45\textwidth]{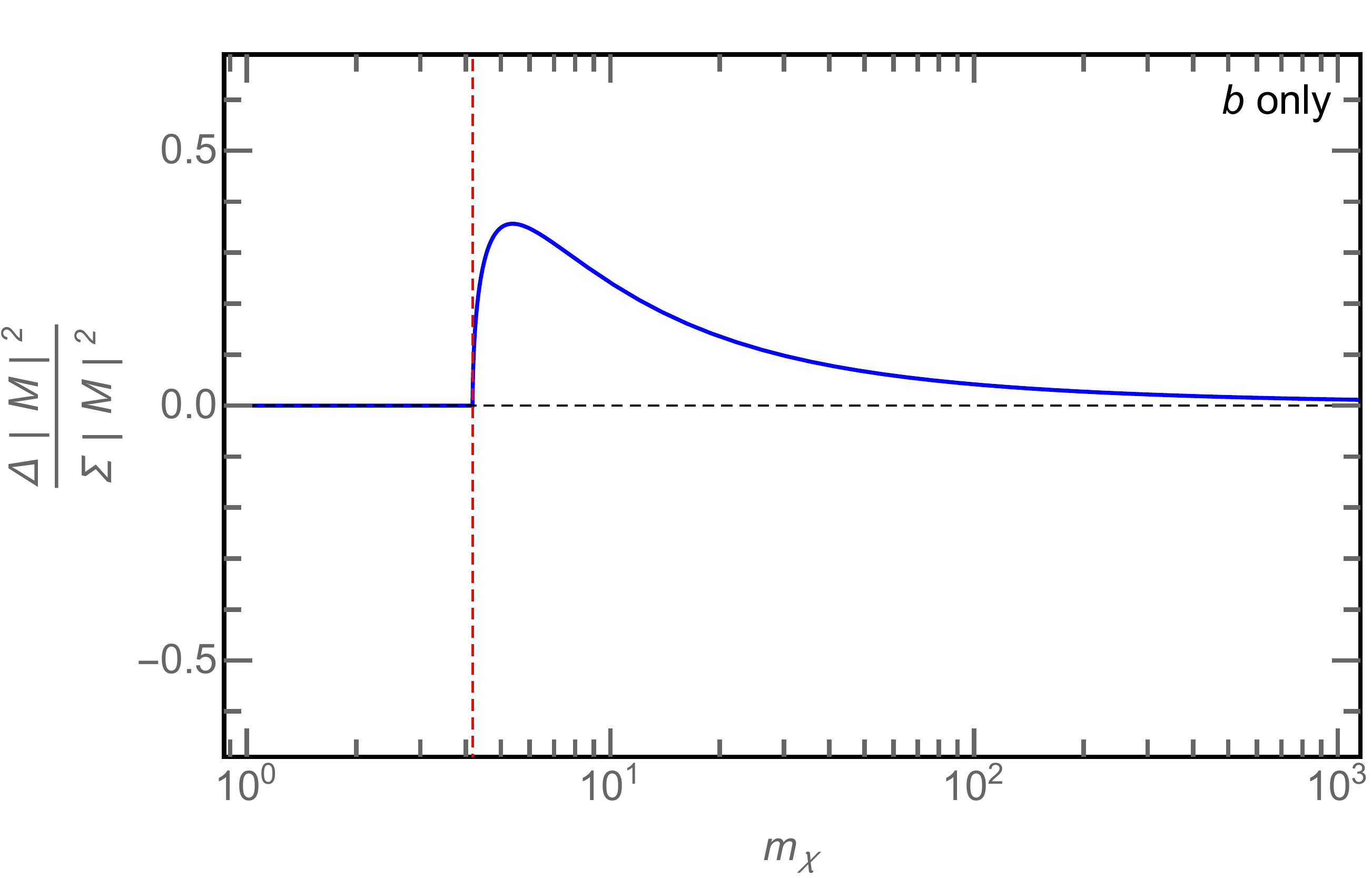} 
\includegraphics[width=0.45\textwidth]{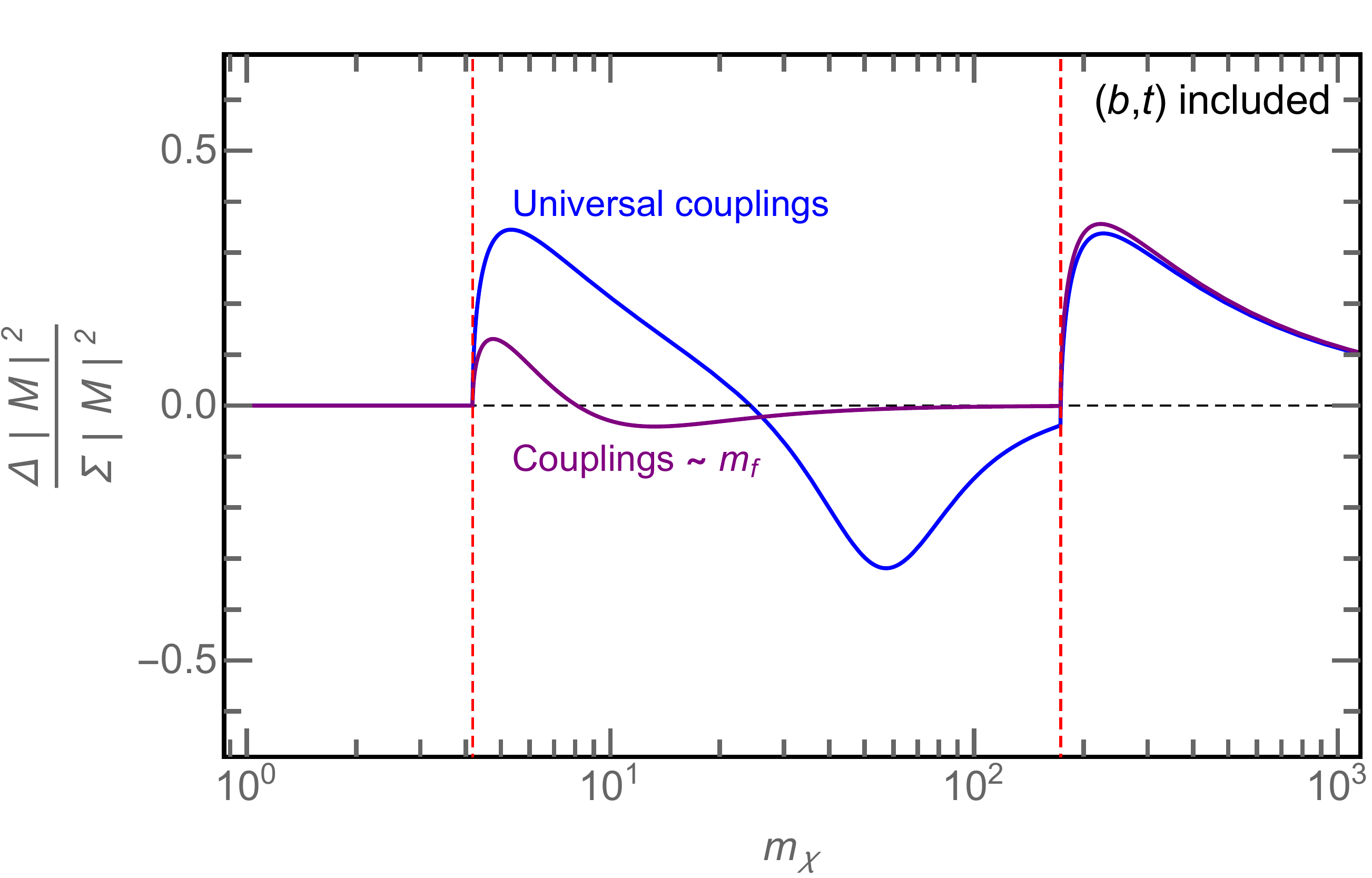} 
\centering
\includegraphics[width=0.45\textwidth]{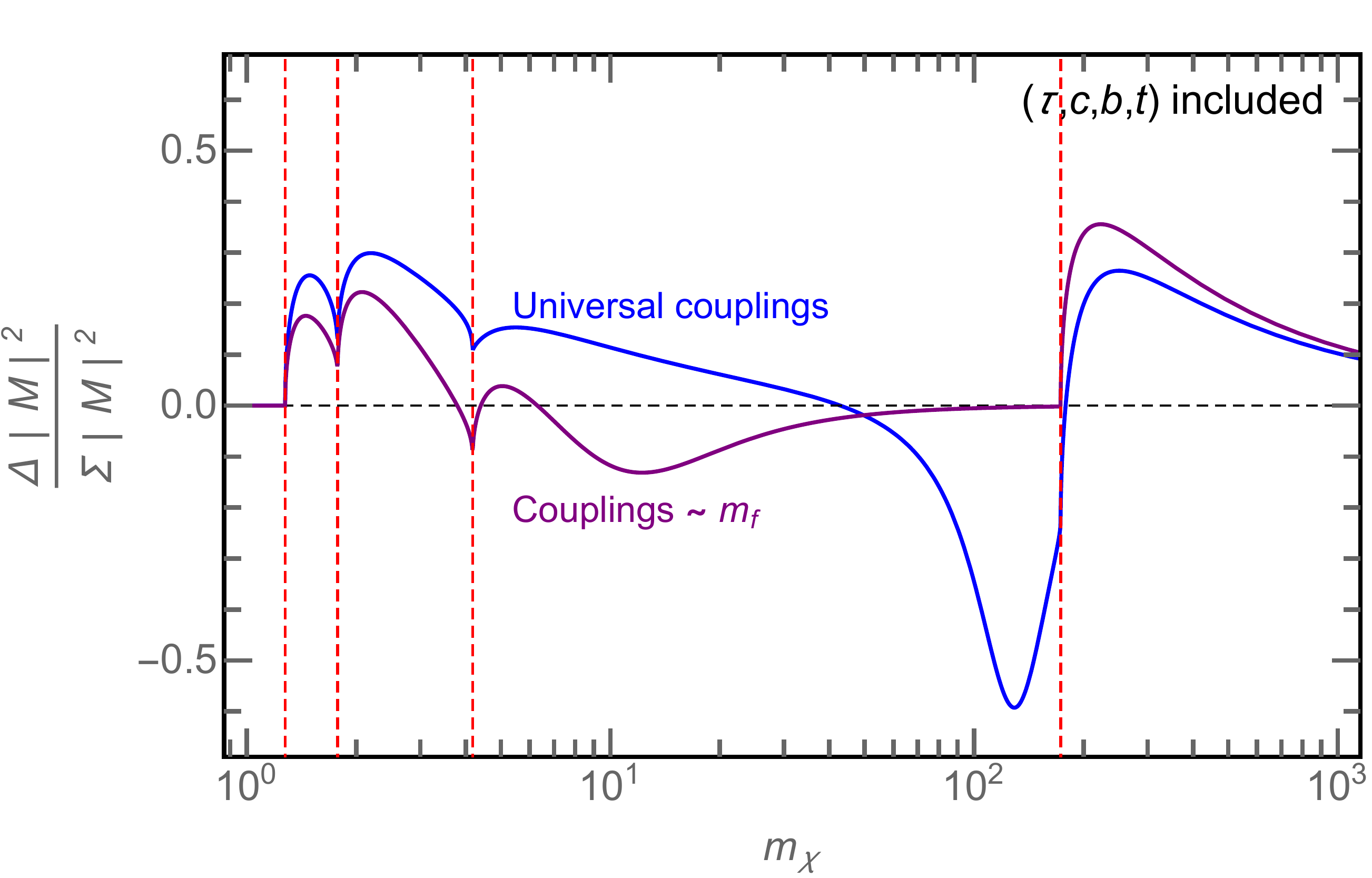} 
\caption{
The polarization asymmetry in the diphoton state for scalar DM with $b$-quark only~(top left), ($b$, $t$)~(top right)
and ($\tau$, $c$, $b$, $t$)~(bottom) contributions. The blue lines correspond to universal couplings $C^S_f=C^P_f$ for all fermions involved, while the purple line assumes couplings scale with the fermion mass:  $C^S_f=C^P_f \sim m_f/m_\chi$. See text for details.}
\label{fig:gaga}
\end{figure}

First, the polarization asymmetry versus DM mass $m_\chi$ for scalar DM with the diphoton final state is shown in Fig.~\ref{fig:gaga}.
The $y$-axis is the polarization asymmetry normalized to the total amplitude squared:
 \begin{align}
\frac{ \vert \sum_f \mathcal  M_f (+, +)\vert^2 - \vert \sum_f \mathcal M_f (-, -)\vert^2}{\vert \sum_f \mathcal  M_f (+, +)\vert^2 + \vert \sum_f \mathcal M_f (-, -)\vert^2} \, ,
 \end{align}
where we consider $b$-quark only~(top left panel), ($b$, $t$)~(top right panel)
and ($\tau$, $c$, $b$, $t$)~(bottom panel) contributions. The blue lines refer to universal couplings $C^S_f=C^P_f$ for all fermions involved, while the purple lines assume the couplings are proportional to the internal fermion mass:  $C^S_f=C^P_f \sim m_f/m_\chi$. The vertical red dashed lines indicate the masses of fermions involved.

From the top left panel, it is clear that polarization asymmetry exists when the internal fermion is on-shell for $m_\chi > m_b$.
In the top right panel,
the blue line exhibits the aforementioned interference effect between heavy-light fermions that can be important for $m_t \geq m_\chi \geq m_b$.
In contrast, the purple line does not feature a significant interplay between the quarks because the  contributions from $b$ 
are suppressed by the coupling for $m_\chi \gg m_b$, leading to a small  interference. 
Finally, the bottom panel shows more complicated interference features if more fermions participate in the processes.

\begin{figure}
\centering
\includegraphics[width=0.45\textwidth]{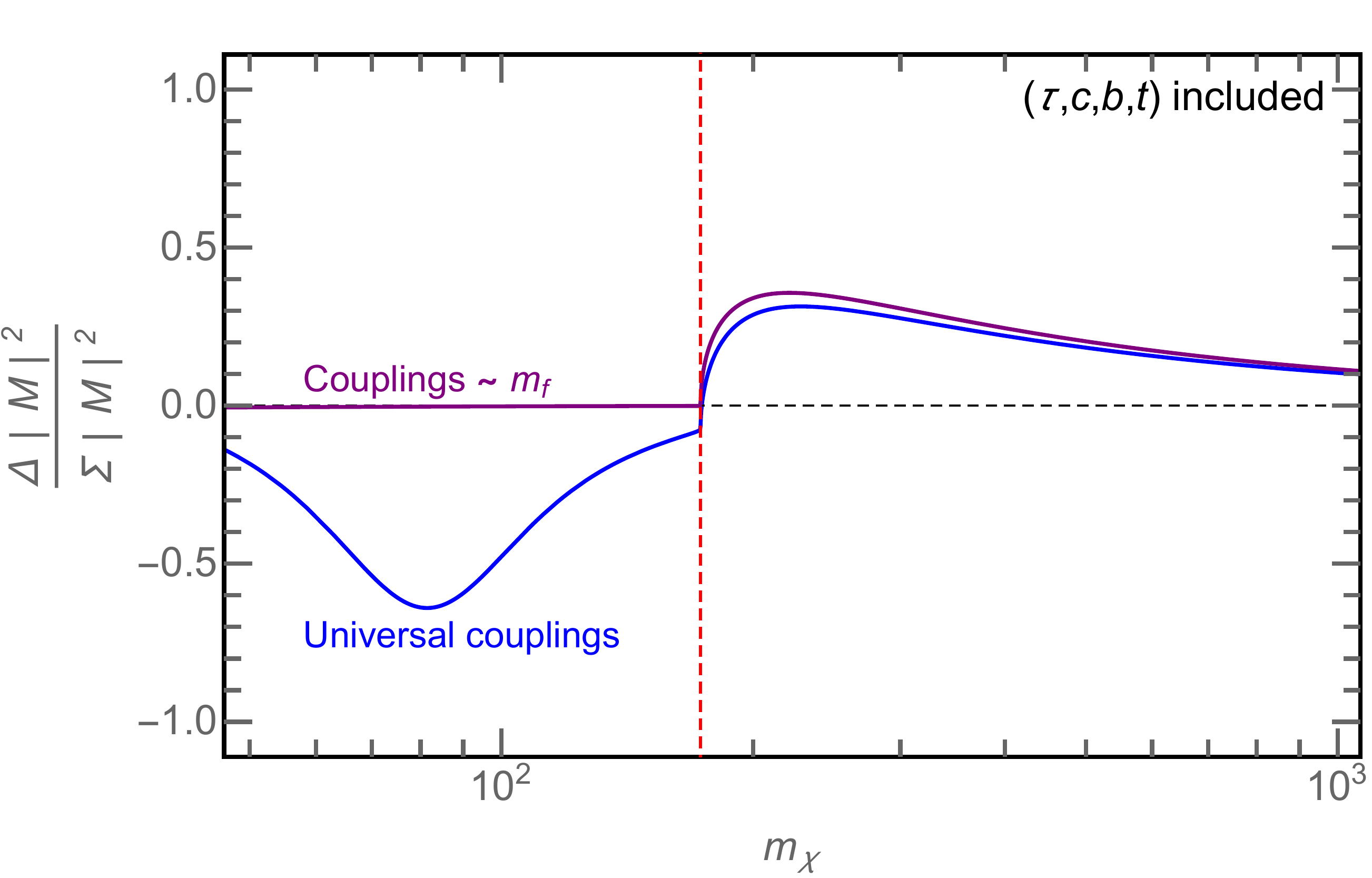} 
\includegraphics[width=0.45\textwidth]{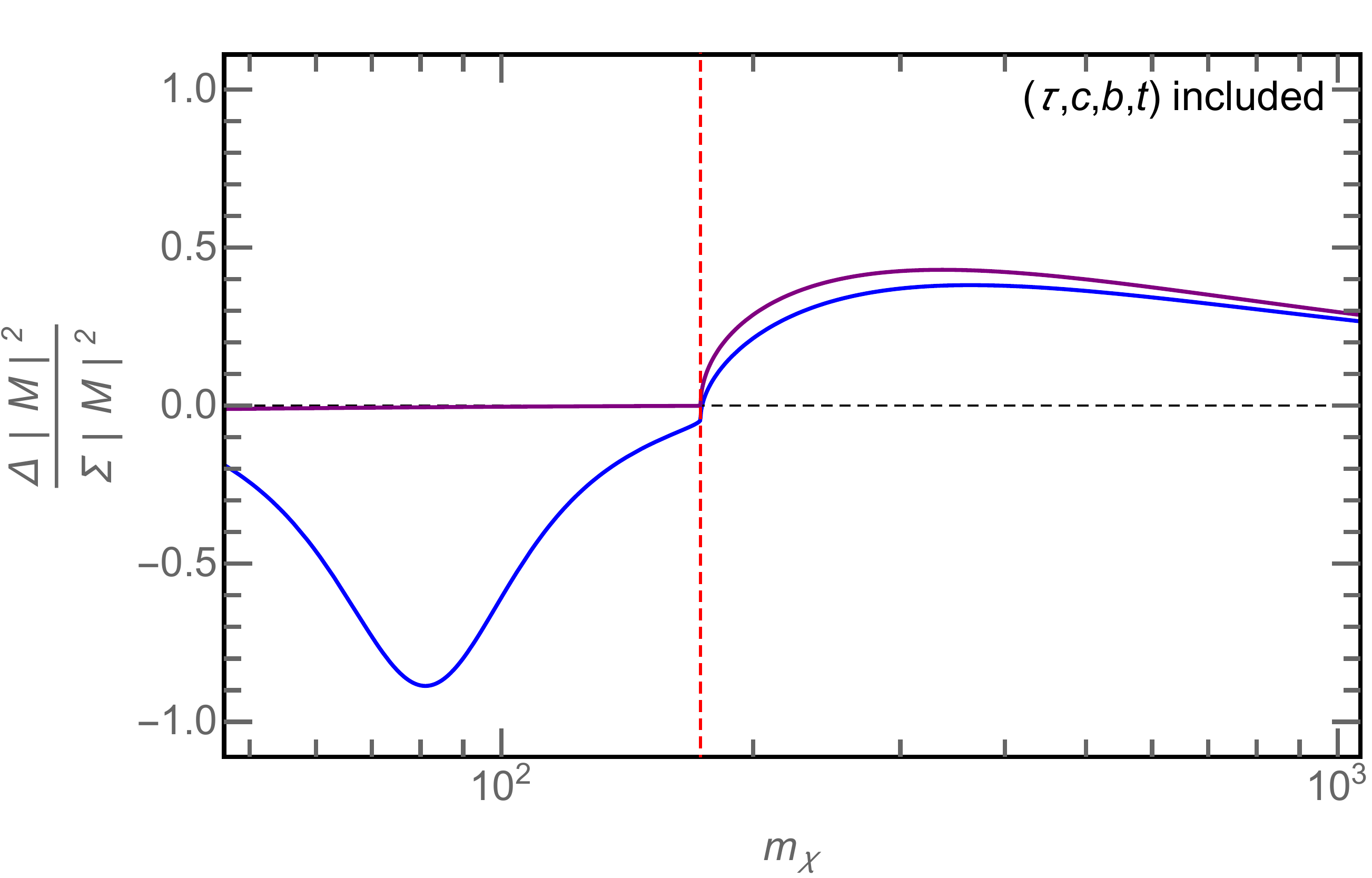}
\caption{Similar to Fig.~\ref{fig:gaga} but for the $Z\ga$ final state.
 Left: Circular polarization asymmetry for scalar DM with $\mathcal L^S$. 
 Right: Circular polarization asymmetry for fermion DM with the tensor operator $\mathcal{L}_3^D$. }
\label{fig:gaZ}
\end{figure}

Next, we display results for the $Z\ga$ final state which, unlike the diphoton channel, can actually generate asymmetry in the 
case of the tensor operator $\mathcal{L}_3^D$.
Note that as the $Z$ boson will eventually decay into SM particles, we sum over all $Z$ polarizations. Therefore, the asymmetry is defined as 
 \begin{align}
\frac{ \vert \sum_{f,Z_{\rm pol}} \mathcal  M_f (Z_{\rm pol} , +)\vert^2 - \vert \sum_{f,Z_{\rm pol}} \mathcal  M_f (Z_{\rm pol} , -)\vert^2}{\vert \sum_{f,Z_{\rm pol}} \mathcal  M_f (Z_{\rm pol} , +)\vert^2 + \vert \sum_{f,Z_{\rm pol}} \mathcal  M_f (Z_{\rm pol} , -)\vert^2} \, .
 \end{align}
The left panel of Fig.~\ref{fig:gaZ} corresponds to the scalar DM with $\mathcal L^S$,
while the right panel presents fermion DM with $\mathcal{L}_3^D$, both with ($\tau$, $c$, $b$, $t$) included in the loop.
As above, we assume a universal coupling $\widetilde C_{\chi f} = (1+i)/\sqrt{2}$~(blue line) and $\widetilde C_{\chi f} = \frac{m_f}{m_\chi}(1+i)/\sqrt{2}$~(purple).
For simplicity, we confine ourselves to the on-shell $Z$ in the final state such that $m_\chi \geq m_Z/2$. 
As can been seen from the plots, the interference effect between the heavy-light fermions is more significant in this case.
Different operators and coupling choices behave quite similarly with $\mathcal{L}_3^D$ having much larger $b$-quark contributions and hence stronger interference effects for $m_\chi \gtrsim m_Z/2$ in the presence of the universal coupling.  

We conclude this section by showing the ratio of fluxes from the loop-induced $\chi \chi \to \ga \ga$ and continuous $\ga$-ray spectrum from the final state radiation~(FSR)
of tree-level DM annihilation processes $\chi \chi \to \bar{f}f \ga$.
The ratio of DM-origin photon numbers in the energy bin of $[m_\chi (1-\eps), m_\chi (1  + \eps)]$~($\eps$: energy resolution of an experiment of interest) between the discrete lines and total contribution reads
\begin{align} 
\frac{N_\ga^{\text{line}}}{N_\ga^{\text{total}}} = \frac{ 2 \lan \sigma v \ran_{\ga\ga} \, 0.68 }{2 \lan \sigma v \ran_{\ga\ga} \, 0.68
+ \lan \sigma v \ran_{\bar{f}f} \De N_\ga
}  \, ,
\label{eq:degree}
\end{align}
where $0.68$ is the probability of the photon line being reconstructed with the energy bin $[m_\chi (1-\eps), m_\chi (1  + \eps)]$, and the
prefactor $2$ comes from the fact that there are two photons in the final state at each DM annihilation.
The symbol $\De N_\ga $ is the number of photons within the energy bin, given a DM annihilation
\begin{align}
\De N_\ga = \int_{m_\chi (1-\eps)}^{m_\chi  (1+ \eps)} \frac{d \Ga }{d E_{\ga}} d E_{\ga} \, ,
\end{align}
where $ \frac{d \Ga }{d E_{\ga}}$ is the FSR photon energy distribution~(vanishing if $E_\ga > m_\chi$) and obtained from {\tt PPPC4DMID}~\cite{Cirelli:2010xx,Ciafaloni:2010ti}.

\begin{figure}
\centering
\includegraphics[width=0.45\textwidth]{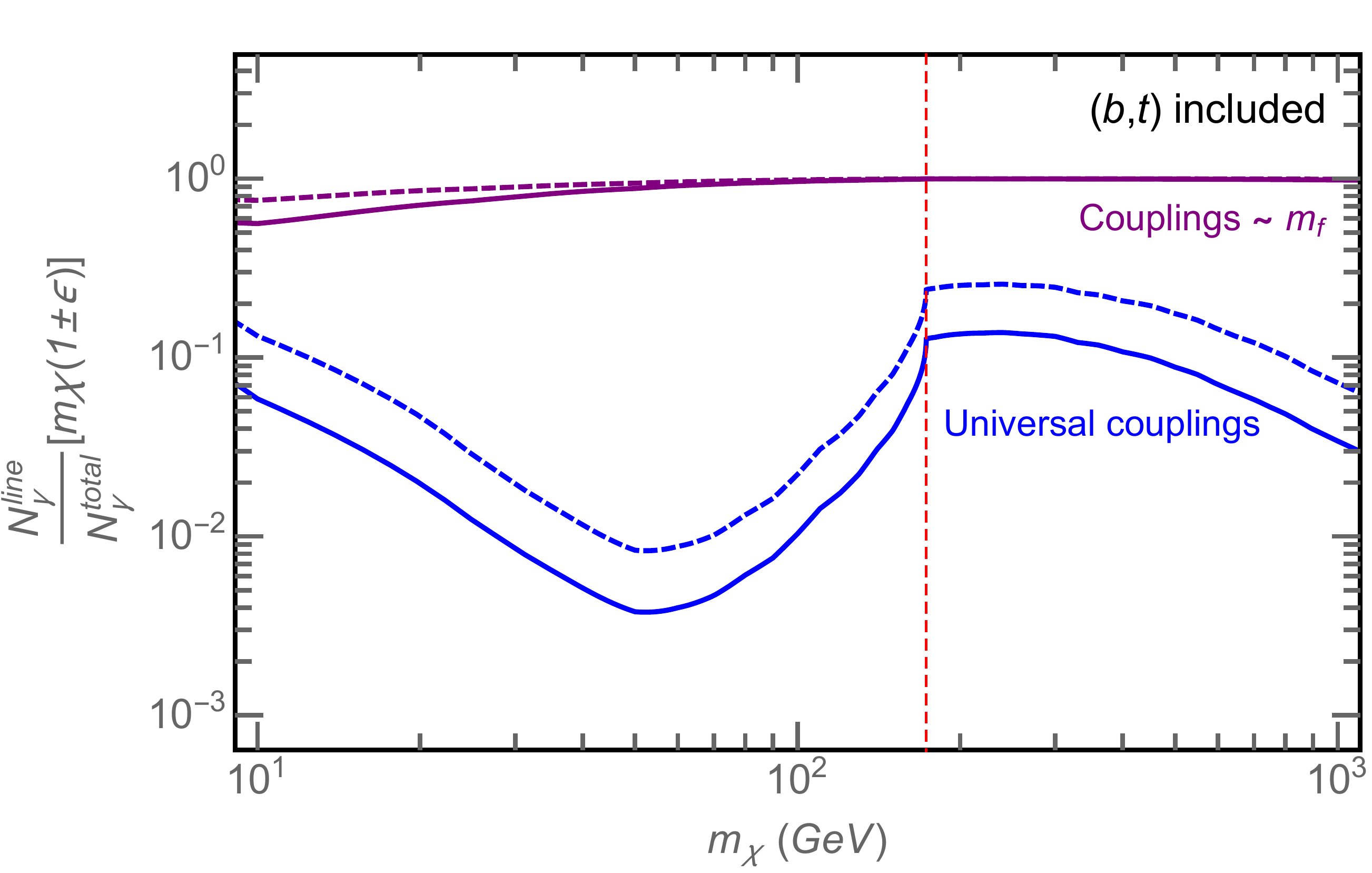} \;\;\;\;
\includegraphics[width=0.45\textwidth]{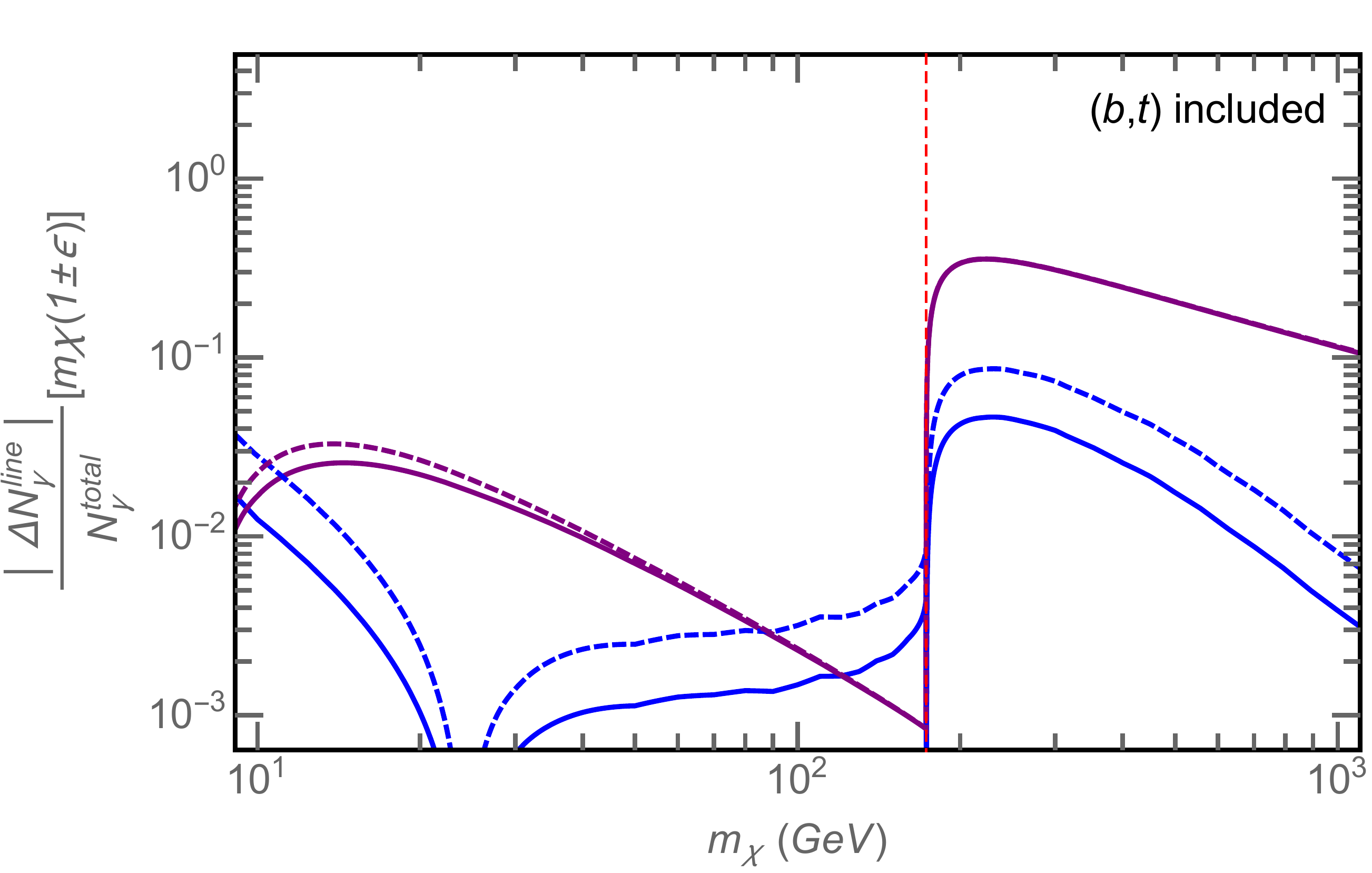} 
\caption{Left: The number ratio of discrete photons to discrete plus continuous ones. 
Right: The number ratio of polarized photons to the total DM-origin photons.
The blue lines correspond to universal couplings while the purple lines indicate
couplings proportional to the mass of fermions in loops. We assume the energy resolution to be $5 \%$~(dashed) and $10 \%$~(solid). See the text for more details.}
\label{fig:lin_con}
\end{figure}

In the left panel of Fig.~\ref{fig:lin_con}, including $b$ and $t$-quarks only 
we have shown the ratio of Eq.~\eqref{eq:degree} for universal couplings~(blue) with $C^S_f=C^P_f$,
and couplings proportional to the mass of fermions~(purple) with $C^S_f=C^P_f \sim m_f/m_\chi$, for scalar DM.
Note that for $m_\chi \geq m_t$, the annihilation channel $\chi \chi \to \bar{t}t$ is open.
We assume the energy resolution to be $10 \%$~(solid lines) and $5 \%$~(dashed lines). It is clear that with a better resolution, the discrete component becomes relatively larger as the decreasing bin width reduces the continuous component.
Moreover, although $\chi \chi \to \ga \ga$ is loop suppressed, the ratio can still be sizable since the FSR photon spectrum diminishes in limit of $E_\ga \to m_\chi$.
Finally, for couplings $ \sim m_f$ the contributions to photon lines from the $t$-quark loop is much more important than $b$-quark, leading to prominent line signals.
This scenario mimics the SM Higgs diphoton decay, where the fermion contributions are dominated by the top quark. 
    
In the right panel of Fig.~\ref{fig:lin_con}, we show a similar plot but replacing the numerator of the ratio~(discrete photon number) by the absolute value of the polarized photon number,
{\it i.e.}, $\De N^{\text{line}}_\ga \equiv \vert N_\ga(+,+) - N_\ga(-,-) \vert$. The asymmetry can be pronounced for $m_\chi \gtrsim m_t$ especially for the case of couplings proportional to masses.
It can be understood from the top right panel of Fig.~\ref{fig:gaga} that the asymmetry is sizable when $m_\chi \gtrsim m_t$ and from the fact that  the line component dominates for large $m_\chi$ as displayed in the left panel of Fig.~\ref{fig:lin_con}.

The total differential DM-origin photon flux by including all annihilation channels denoted by $i$ from the Galactic Center is
\begin{align}
\frac{d^2 \Phi^{\rm total}_{\chi} }{d \Omega \, dE_\ga} = \frac{1}{2} \frac{r_\odot}{4 \pi}
\lee \frac{\rho_\odot}{m_{\chi}} \rii^2 J \sum_i \lan \sigma v \ran_i \frac{d \Ga_i}{dE_\ga} \, ,
\end{align}
where $d \Omega$ is the solid angle, $r_\odot$ is the distance from the Galactic Center to the Sun, 
$\rho_\odot$ is the local DM density,
and $J$ is the $J$-factor which is the integration of DM contributions along the line of sight.
If both DM particle and antiparticle are present, an extra $1/2$ is needed.
Assume the astrophysical $\ga$-ray background be unpolarized and its differential flux be denoted by
\begin{equation}
\frac{d^2 \Phi_{\text{bkg}}}{d \Omega \, dE_\ga} .
\end{equation}
Then, the number of background photons that contributes to the $N_\ga^{\text{total}}$ in Eq.~(\ref{eq:degree}) is given by
\begin{equation}
N_\ga^{\text{bkg}} = \frac{d\Phi_{\text{bkg}}}{d \Omega} \left[ \frac{1}{2} \frac{r_\odot}{4 \pi}
\lee \frac{\rho_\odot}{m_{\chi}} \rii^2 J \right]^{-1},
\end{equation}
where
\begin{align*}
\frac{d \Phi_{\text{bkg}} }{d \Omega} = \int_{m_\chi (1-\eps)}^{m_\chi (1+\eps)} \frac{d^2 \Phi_{\text{bkg}}}{d \Omega \, dE_\ga} \, dE_\ga \, .
\end{align*}
Thus the degree of circular polarization will be lowered by the unpolarized $\ga$-ray background. That can be remedied by increasing the energy resolution of the $\ga$-ray polarimetry to capture the polarized line photons.

\section{Prospects for detecting a net circular polarization}\label{sec:prospects}

The azimuthal angle of the plane of production of an electron-positron pair created in a $\ga$-ray detector provides a way of measuring linear polarization of incoming $\ga$ rays.  It has been demonstrated that the use of an active target consisting of a time-projection chamber enables the measurement of the linear polarization with an excellent effective polarization asymmetry~\cite{Gros:2017wyj}. The current $\ga$-ray detectors are not designed primarily for polarization measurement. Instruments sensitive to linear polarization will be employed in future $\ga$-ray experiments such as AdEPT, HARPO, ASTROGAM, and AMEGO, with the minimum detectable polarization (MDP) from a few percents up to $20\%$~\cite{Knodlseder:2016pey, Moiseev:2017mxg}. In principle, the measurement of bremsstrahlung asymmetry of secondary electrons produced in Compton scattering off a magnetized or unpolarized target can be used to determine the circular polarization of incoming $\ga$ rays. However, no efficient methods using non-Compton scattering techniques for measuring $\ga$-ray circular polarization have been developed to date. Improved or even new techniques for $\ga$-ray circular polarimetry are yet to be explored.

In Ref.~\cite{Elagin:2017cgu}, the authors have discussed the possibility of detecting the circular polarization asymmetry of the $\ga$-ray flux in future $\ga$-ray polarimetry experiments. Optimistically, to produce one useful event that can be used in the secondary asymmetry measurement would need about $10^3$ photons. The total number of useful events required to measure an asymmetry at one sigma level can be estimated by $N_{\text{useful}}\sim (AP_\gamma)^{-2}$, where $A$ is the asymmetry generated by a polarized photon and $P_\gamma$ is the fraction of circular polarization. In the present work, $P_\gamma$ can reach $0.4$ at $E\sim 200\, {\rm GeV}$. Assuming that $A\sim  0.1$, to detect a $40\%$ polarized DM signal, we must collect a number of $\ga$ photons roughly equal to $10^3 N_{\text{ useful}} \sim 6\times 10^5$. 

The possible $\ga$-ray excess from the Galactic Center has been suggested by the Fermi-LAT observations~\cite{Calore:2014xka}. The $\ga$-ray flux at $E\sim 200\, {\rm GeV}$ can be fitted by
\begin{equation}
\frac{d^2 \Phi_{\text{excess}}}{d \Omega \, dE_\ga} \sim 10^{-7} \left(\frac{\rm GeV}{E_\ga}\right)^2 {\rm GeV^{-1}cm^{-2}s^{-1}sr^{-1}}.
\end{equation}
Assume the excess $\ga$-ray flux be dominated by the DM signal. Then, the number of $\ga$ photons that go through a detector is given by
\begin{equation}
\label{eq:val_ph}
\frac{d^2 \Phi_{\text{excess}}}{d \Omega \, dE_\ga}\, 2\eps E_\ga\, I_{\rm exp}\, \Delta\Omega,
\end{equation}
where $I_{\rm exp}$ is the detector exposure and $\Delta\Omega$ is the subtended solid angle of the Galactic Center. By taking 
$E=200\, {\rm GeV}$, $\eps=0.1$, $I_{\rm exp}=5000\,{\rm cm^2\,yr}$, and $\Delta\Omega=0.18$, we find that the number of $\ga$ photons is about 3, which is far below the required number. Note that for lighter DM, the increase on the incoming photon flux~(Eq.~\eqref{eq:val_ph}) is unfortunately offset by the decrease of induced polarization asymmetry
as shown in the right panel of Fig.~\ref{fig:lin_con}, leading to the same conclusion.
Future $\ga$-ray polarimetry experiments would need to largely improve the asymmetry measurement and the number of useful events. Otherwise, it seems that new technologies for detecting a net circular polarization in photons should be explored.

\section{Conclusions}\label{sec:conclusions}

We have studied the possibility for a net circular polarization of the $\ga$ rays coming from dark matter annihilations. We have considered the effective couplings between the fermions in the Standard Model and neutral scalar, Dirac, and Majorana dark matter, which annihilate into monochromatic diphoton and $Z$-photon final states. 
The circular polarization asymmetry in the diphoton and $Z$-photon states for the scalar dark matter can be substantial~(even up to nearly $90\%$ for the $Z$-photon channel), provided that $P$ and $CP$ symmetries are violated in the couplings and internal fermions are on-shell. Given the energy resolution of a $\ga$-ray detector at $5-10\%$ level, the degree of circular polarization at the dark matter mass threshold can reach $10-40\%$ for the dark-matter induced $\ga$-ray flux coming from the Galactic Center. The unknown astrophysical $\ga$-ray background would obscure the detectability. However, we can make use of the line spectrum of the $\ga$-ray flux from dark matter annihilations to single out the polarization signals from the background, if unpolarized, and the continuum photons resulting from annihilating final-state interactions.

\section*{Acknowledgments}
We would like to thank Denis Bernard for a private communication.
This work was supported in part by the Ministry of Science and Technology (MoST) of Taiwan under
grant numbers 107-2119-M-001-030 (KWN) and 107-2119-M-001-033 (TCY).
WCH was supported by the Independent Research Fund Denmark, grant number 
DFF 6108-00623. The CP3-Origins centre is partially funded by the Danish National Research Foundation, 
grant number DNRF90. This work was partially performed at the Aspen Center for Physics, which is supported by National Science Foundation grant PHY-1607611.

\newpage

\appendix

\section{$Z'$ toy model}
\label{sec:UV_mod}

\begin{figure}
\centering
\includegraphics[width=0.4\textwidth]{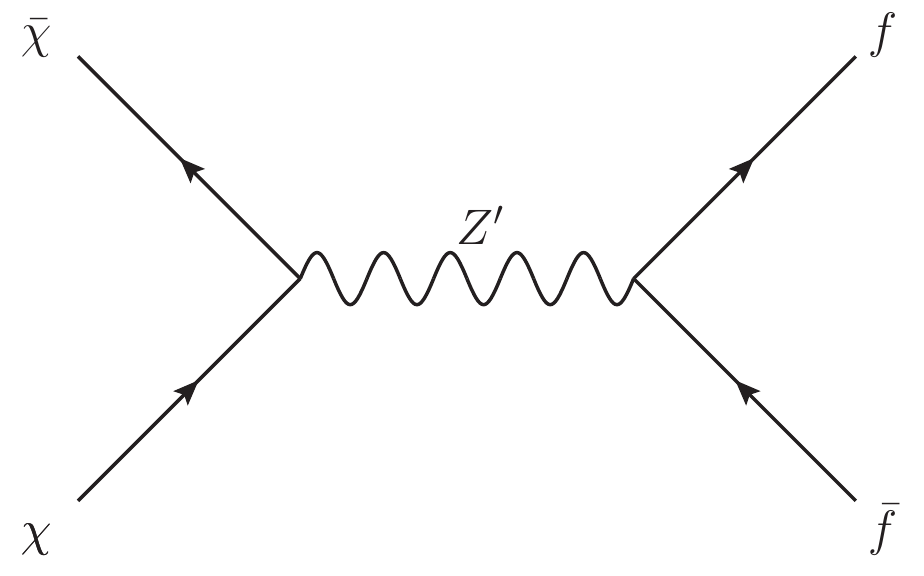} \;\; \;\; \;
\includegraphics[width=0.53\textwidth]{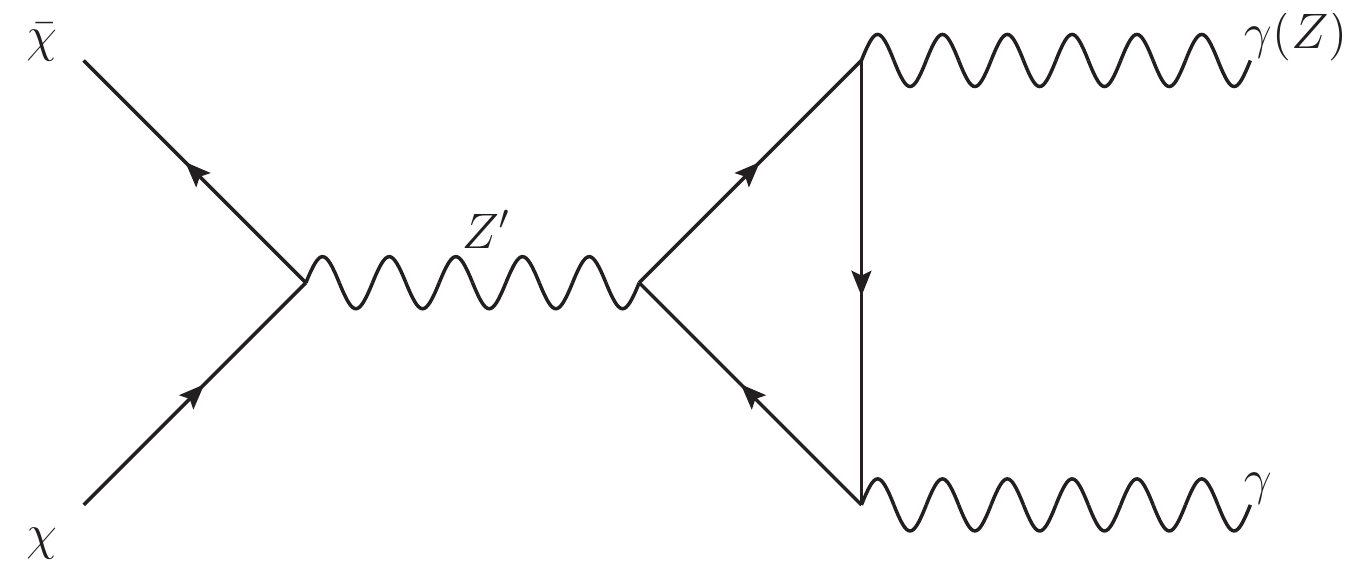}
\caption{Left: DM annihilates into SM fermions.
Right: DM annihilates into $\gamma\gamma$ or $Z\gamma$ via loop dressing by SM charged fermions.}
\label{fig:U1_chif}
\end{figure}

We here show that a toy model of an abelian gauge symmetry $U(1)'$ with the corresponding $Z'$ gauge boson will generate the same result as predicted
by the effective approach. Assuming both DM particles and SM fermions are charged under the $U(1)'$, thus DM can annihilate into SM fermions via the $Z'$ exchange
as shown in the left panel of Fig.~\ref{fig:U1_chif}, as well as the loop-induced $Z\ga$ and $\ga\ga$ channels in the right panel of Fig.~\ref{fig:U1_chif}.

Depending on the $U(1)'$ charge assignment on $\chi$ and $f_{L,R}$, the coupling strength can be different for the left-handed and right-handed fields;
for instance, in the limit of $m_{Z'} \gg 2 m_\chi$, one has for Eq.~\eqref{eq:effL_D2}
\begin{align}
C^L_\chi = C^R_\chi = C^L_f = C^R_\chi =1 
\;\;\; \text{and} \;\;\; 
\frac{ \left( 4 \pi \right)^2}{ \Lambda^2} = \frac{1}{ m^2_{Z'}} \; .
\end{align}

Note that the loop structure in the UV model is exactly the same as those in the effective approach.
As a consequence, one should obtain the same result from the UV model and effective approach. It alludes to the main point in this Appendix that our results only apply to
the specific one-loop structure which contains either SM or new fermions only and also the mediator~($Z'$ in this case) has to be heavier than twice the DM mass.


\newpage


\end{document}